\documentclass[a4paper,11pt]{article}
\pdfoutput=1 

\usepackage{jcappub} 
\usepackage{amsthm}

\usepackage[T1]{fontenc} 
\usepackage{mathtools}

\usepackage{enumitem}

\usepackage{aas_macros}
\usepackage{amssymb}
\usepackage{multirow}
\usepackage{cleveref}
\usepackage{siunitx}
\usepackage{enumerate}
\usepackage[dvipsnames]{xcolor}
\usepackage{float}

\usepackage{tikz-feynman}
\usepackage{afterpage}

\usepackage{psfrag}
\usepackage{color}
\usepackage{graphicx}
\usepackage{upgreek}
\usepackage{feynmp}
\DeclareMathAlphabet{\mathpzc}{OT1}{pzc}{m}{it}

\usepackage[mathscr]{euscript}
\usepackage{bbm}

\usepackage{tikz}
\usetikzlibrary{trees}
\usetikzlibrary{decorations.pathmorphing}
\usetikzlibrary{decorations.markings}
\tikzset{
    photon/.style={decorate, line width=0.15mm, decoration={snake,amplitude=3pt,segment length=8pt}, draw=black},
    wino/.style={draw=redwine},    
    fermion/.style={draw=black, line width=0.2mm, postaction={decorate},
        decoration={markings,mark=at position .55 with {\arrow[draw=black,scale=2,#1]{>}}}},
    scalar/.style={draw=black, dashed,postaction={decorate},
        decoration={markings,mark=at position .55 with {\arrow[draw=black,scale=2,#1]{>}}}},
    scalarline/.style={draw=black, postaction={decorate},
        decoration={markings,mark=at position .55 with {\arrow[draw=black,scale=2,#1]{>}}}},
    scalarline2/.style={draw=black, postaction={decorate} },
    scalar2/.style={draw=black, dashed,postaction={decorate}},
    gluon/.style={decorate, draw=black,
        decoration={coil,amplitude=3pt, segment length=4pt}},
    graviton/.style={decorate, draw=black,
        decoration={zigzag,amplitude=3pt, segment length=4pt}}
}
\tikzstyle{blob}=[circle,
                              minimum size=20,
                              draw=black!80,
                              fill=black!80]   
\tikzstyle{redblob}=[circle,
                              thick,
                              minimum size=0.4cm,
                              draw=red!80,
                              fill=red!60]

\allowdisplaybreaks[4]

\definecolor{darkgreen}{rgb}{0,0.5,0}

\newcommand{\beq}{\begin{eqnarray}}
\newcommand{\eeq}{\end{eqnarray}}

\newcommand{\bseq}{\begin{subequations}}
\newcommand{\eseq}{\end{subequations}}
\newcommand{\be}{\begin{equation}}
\newcommand{\ee}{\end{equation}}

\def\d{\partial}
\def\e{{\rm e}}
\def\d{\partial}

\renewcommand{\Im}{\mathop{\rm Im}\nolimits}
\renewcommand{\Re}{\mathop{\rm Re}\nolimits}

\newcommand{\ve}{\varepsilon}

\renewcommand{\k}{\mathbf{k}}
\newcommand{\x}{\mathbf{x}}
\newcommand{\E}{\mathcal{E}}

\newcommand{\beqa}{\begin{eqnarray}}
\newcommand{\eeqa}{\end{eqnarray}}

\newcommand\vf{\varphi}

\def\d{\partial}

\title{
Boson Star Normal Modes}

\author[a,b,c]{James Hung-Hsu Chan,}
\author[d,e]{Sergey~Sibiryakov,}
\author[f]{and Wei~Xue}

\affiliation[a]{
Department of Astrophysics, American Museum of Natural History,\\
Central Park West and 79th Street, NY 10024-5192, USA} 
\affiliation[b]{
Department of Physics and Astronomy, Lehman College of the CUNY,\\
Bronx, NY 10468, USA }
\affiliation[c]{
Institute of Physics, Laboratory of Astrophysique,
\' Ecole Polytechnique F\' ed\' erale\\ de Lausanne (EPFL),
Observatoire de Sauverny, 1290 Versoix, Switzerland}
\affiliation[d]{Department of Physics \& Astronomy, McMaster
University,\\
Hamilton, Ontario, L8S 4M1, Canada} 
\affiliation[e]{Perimeter Institute for Theoretical Physics, Waterloo,
 Ontario, N2L 2Y5, Canada}
\affiliation[f]{Department of Physics, University of Florida,
  Gainesville, FL 32611, USA} 

\emailAdd{jchan@amnh.org}
\emailAdd{ssibiryakov@perimeterinstitute.ca}
\emailAdd{weixue@ufl.edu}

\abstract{
Boson stars are gravitationally bound objects that arise in ultralight
dark matter models and form in the centers of galactic halos or axion
miniclusters. We systematically study the excitations of a boson star,
taking into account the mixing between positive and negative
frequencies introduced by gravity. We show that the spectrum contains
zero-energy modes in the monopole and dipole sectors resulting from
spontaneous symmetry breaking by the boson star background. We analyze
the general properties of the eigenmodes and derive their
orthogonality and completeness conditions which have non-standard form
due to the positive-negative frequency mixing. The eigenvalue problem
is solved numerically for the first few energy levels in different
multipole sectors and the results are compared to the solutions of the
Schr\"odinger equation in fixed boson star gravitational potential. The two
solutions differ significantly for the lowest modes, but get close for
higher levels. We further confirm the normal mode spectrum in 3D wave
simulations where we inject perturbations with different multipoles. 
As an application of the normal mode solutions, we compute the matrix
element entering the evaporation rate of a boson star immersed in a
hot axion gas. The computation combines the use of exact wavefunctions
for the low-lying bound states and of the Schr\"odinger approximation
for the high-energy excitations.    
}

\begin{document}
\maketitle
\flushbottom

%
%
%

\section{Introduction}
\label{sec:intro}

Ultralight bosonic dark matter~(ULDM) is an attractive dark matter (DM) 
candidate. Well-motivated examples of ULDM are the QCD axion 
\cite{Weinberg:1977ma,Wilczek:1977pj,Shifman:1979if,Kim:1979if,Zhitnitsky:1980tq,Dine:1981rt,Preskill:1982cy,Abbott:1982af,Dine:1982ah}  
and axion-like particles
\cite{Arvanitaki:2009fg,Marsh:2015xka,Hui:2016ltb,Hui:2021tkt}. 
The QCD axion was
originally proposed to solve the strong CP problem
\cite{Peccei:1977hh,Peccei:1977ur}, and  
the axion-like particles 
arise from the context of beyond the Standard Model scenarios and
string theory \cite{Svrcek:2006yi,Arvanitaki:2009fg,Hui:2016ltb}. 
Both can naturally produce the observed DM abundance from the early
universe evolution.  

Compared with other DM candidates, ULDM shares a similar quality in
explaining the DM observations in the large-scale structure, 
but has novel properties on a small scale
that is about its de Broglie wavelength. 
As a distinct feature of ULDM, coherent soliton solutions, known as
{\it axion solitons} or {\it boson stars} \cite{Ruffini:1969qy},  
exist due to the interplay between the self-gravity of the axion field
and its wave properies.  
All the axions in a boson star are in the same ground state supported
by the gravitational potential. 
Because the soliton solutions are energetically favorable, they are
formed by the kinematic relaxation of axions through gravitational
interactions  
\cite{Levkov:2018kau} or other mechanisms.
The formation of these objects in the central region of axion dark
matter halos (also known as the axion mini-clusters  
\cite{Hogan:1988mp,Kolb:1993zz} for the QCD axion)
is confirmed by several numerical simulations
in the cosmological setting
\cite{Schive:2014dra,Veltmaat:2018,Mina:2020eik,May:2021wwp}, kinetic
relaxation region  
\cite{Levkov:2018kau,Chan:2022bkz} or by considering the collision of
several seed solitons
\cite{Schive:2014hza,Schwabe:2016rze,Mocz:2017wlg}. 
Even if the soliton is artificially removed from the halo, evolution
will restore it back \cite{Yavetz:2021pbc}. 
The size of the formed solitons is about the de Broglie wavelength of
the surrounding ULDM.  
The ULDM with mass $m_a\sim 10^{-22}\div 10^{-19}~{\rm eV}$, known as
fuzzy DM \cite{Hu:2000ke},  
has the wavelength of many parsecs, so is the size of soliton
in the center. Tight constraints have been put on fuzzy DM from
Ly-$\alpha$ forest data
\cite{Armengaud:2017nkf,Kobayashi:2017jcf,Rogers:2020ltq} and dynamics
of dwarf galaxies 
\cite{Marsh:2018zyw,Dalal:2022rmp}. 

The observational signatures
from axion solitons open new opportunities to discover or constrain
axion DM. For the fuzzy DM, the presence of solitons 
in the center of galaxies can modify the galaxy rotation curves
\cite{Bar:2018acw,Bar:2021kti}, 
and heat the stars in the central region via oscillations and random
walk \cite{Marsh:2018zyw,Chiang:2021uvt,Chowdhury:2021zik}. 
Also, the presence or absence of axion solitons have important
implications for lensing search \cite{Ellis:2022grh}.  
When considering the self-interactions \cite{Levkov:2016rkk} and
couplings to the photons
\cite{Tkachev:2014dpa,Hertzberg:2018zte,Levkov:2020txo},  
the axion solitons can produce relativistic axions
and radio emissions, respectively. 
The dense axion stars made of inflaton can seed primordial black holes
or produce stochastic high-frequency gravitational wave signals 
\cite{Eggemeier:2021smj}.

All these interesting signatures are based on the fact that 
the existence of axion solitons appears as a generic feature of an axion halo. 
This generic feature is observed through extensive studies on the
formation and evolution of axion solitons  
\cite{Schive:2014dra,Veltmaat:2018,Mina:2020eik,May:2021wwp,Levkov:2018kau,Chan:2022bkz}, 
which rely on the interactions between the solitons and their
environment, the surrounding 
axion waves. Refs. \cite{Schive:2014dra, Schive:2014hza} find an
interesting correlation between the soliton mass and the host halo
mass 
in cosmological numerical simulations, and the relationship is
confirmed in \cite{Veltmaat:2018,Eggemeier:2019jsu}. Although the
soliton-host  
halo relation appears intuitive by equating the virial temperature of
the soliton and the host halo \cite{Chavanis:2019faf}, 
the underlying mechanism remains
unclear.  
It is not reproduced by other simulations in non-cosmological setting
\cite{Schwabe:2016rze,Mocz:2017wlg,Chan:2021bja,Zagorac:2022xic}, and 
more recent cosmological simulations
\cite{Nori:2020jzx,May:2021wwp,Chan:2021bja} observe a larger scatter
of the relation. 
In the kinetic regime, the soliton-host halo relation is disfavored
because a continuous growth of solitons after their formation 
has been observed \cite{Levkov:2018kau}, while
refs.~\cite{Eggemeier:2019jsu,Chen:2020cef} argue that the growth
slows down when the soliton 
becomes heavy enough to heat up the inner part of axion halos. 

In ref.~\cite{Chan:2022bkz}, by combining the analytic approach and
numerical simulations  
we find that, somewhat counter-intuitively, solitons with virial
temperature higher than the temperature of the ambient axion gas
grow by condensation from the gas, whereas solitons colder than a certain
threshold evaporate. Since the temperature of the soliton is
proportional to the square of its mass, the former solitons were
dubbed in \cite{Chan:2022bkz} as `heavy', and the latter ones as `light'.  
From the analytic computations, it is 
evident that the dynamics of solitons arise from the scattering
between a soliton and its surrounding axion waves. 
To explore the dynamics of solitons, 
we need to understand the eigenstates of axion waves in the soliton background.

The complexity of the axion modes occurs because the modes with
positive and negative frequencies with 
respect to the soliton background are coupled together by gravity
through the Schr\"odinger--Poisson (SP) system of equations. 
Two approaches to this problem exist in the literature. 
Refs.~\cite{Yavetz:2021pbc,Zagorac:2021qxq,Zagorac:2022xic} 
neglect the mixing of
modes so that the equation is simplified to a linear  
Schr\"{o}dinger equation in fixed gravitational potential of the
soliton and the halo.
This approximation seems to work well for studies of collective
properties of the axion eigenstates. However, it fails in some
important details --- in particular, it does not yield a dipole
zero-energy mode whose presence is required by the (spontaneously
broken) translation invariance.
On the other hand, 
refs.~\cite{Harrison_2003,Guzman:2004wj,Guzman:2018bmo} find numerically several
low-lying excited states, for which the mode mixing plays an essential
role. 

The goal of the paper is to
connect the two approaches and to clarify whether 
and when we can neglect mode mixing. 
Our analysis follows the lines of
refs.~\cite{Guzman:2004wj,Guzman:2018bmo}, complementing them in
several ways. We lay out a general framework to find all eigenstates
and derive their orthogonality and completeness properties. We find
numerically the energies and wavefunctions of all low-lying sates with
$\ell+n\leq 4$, where $\ell$ is the orbital (multipole) quantum number
and $n\geq 0$ is the level number within the given multipole
sector.\footnote{In the language of atomic physics $\ell+n+1$ would
  correspond to the principal quantum number. For dipole, $\ell=1$, we
go up to $\ell+n=5$.} We confirm that the positive-negative mode mixing
is important for accurate determination of the lowest energy levels
and show how it leads to recovery of the zero-energy dipole mode. On
the other hand, we observe that for high-frequency excitations
corresponding to $\ell+n\geq 4$ the solutions of the
SP system are well approximated by the
eigenfunctions of the Schr\"odinger equation in fixed gravitational
potential. We further improve the approximation by developing a
perturbative expansion for exact eigenvalues around the eigenvalues of
the Schr\"odinger equation. These results are compared with
spectroscopic measurements of a perturbed soliton in
$(3+1)$-dimensional simulations and are found to be in excellent
agreement.   

The complete axion wavefunctions and their high-frequency approximation are
crucial information to deduce the scattering rate of axion waves on
the soliton.  
We illustrate this point on the example of the light soliton
evaporation rate \cite{Chan:2022bkz} which we evaluate by combining
the advantages of the exact and the Schr\"odinger wavefunctions. 

The paper is organized as follows. In \cref{sec:basicEq}, we introduce
the SP equations for soliton background
and perturbations.
We analyze the general properties of the eigenfunctions in
\cref{sec:perturbations}. Then we numerically  
solve the SP equations 
and Schr\"{o}dinger equations to obtain the exact and approximate
wavefunctions in \cref{sec:PertTheory}. Soliton spectroscopy using 
3D wave simulation is presented in
\cref{sec:wavesimulation}. In \cref{sec:scattering} we explore the
interaction between 
the soliton and axion waves and evaluate the evaporation
rate of a light soliton. We conclude in
\cref{sec:conclusion}. Appendix contains some details about our
numerical procedure to solve the eigenvalue problem.

\section{Soliton and its Perturbations}
\label{sec:basicEq}

We consider a non-relativistic complex scalar field $\psi(t, {\bf x})$ only with 
gravitational self-interactions. Its dynamics follows the
Schr\"{o}dinger--Poisson (SP) equations, 
\bseq
\label{eq:SPequation}
\begin{align}
\label{eqSchr}
&i\d_t\psi +\frac{\Delta\psi}{2m}-m\Phi\psi=0\;,\\
\label{eqPois}
&\Delta\Phi=4\pi G m\,|\psi|^2 \;,
\end{align}
\eseq
where $\Delta$ denotes the Laplacian. The density of scalar field is
$\rho (t, {\bf x} ) = m |\psi(t, {\bf x})|^2$, 
and $\Phi$ is the gravitational potential given by the density. 
The system has scaling symmetry 
\begin{equation}
   \psi ( t, {\bf x} ) \to { {\cal S}^2}  \, \psi  (   {{\cal S}^2} \, t ,   {{\cal S}} \, {\bf x})  \, ,
   \quad
   \Phi ( t, {\bf x} ) \to { {\cal S}^2}  \, \Phi (   {{\cal S}^2} \, t ,   {{\cal S}} \, {\bf x}) \, , 
   \label{eq:scaling}
\end{equation}
which can be used to generate a family of solutions from a single
solution. More general scaling transformations given in \cite{Chan:2022bkz}
allow one to eliminate $m$ and $4\pi G$ and work with dimensionless
quantities. However, we prefer to keep
$m$ and $4\pi G$ explicitly in the analytic derivations and only
switch to the units $m=4\pi G=1$ in the numerical calculations.

\begin{figure}[h]
\begin{center}
 \includegraphics[width=0.45\textwidth]{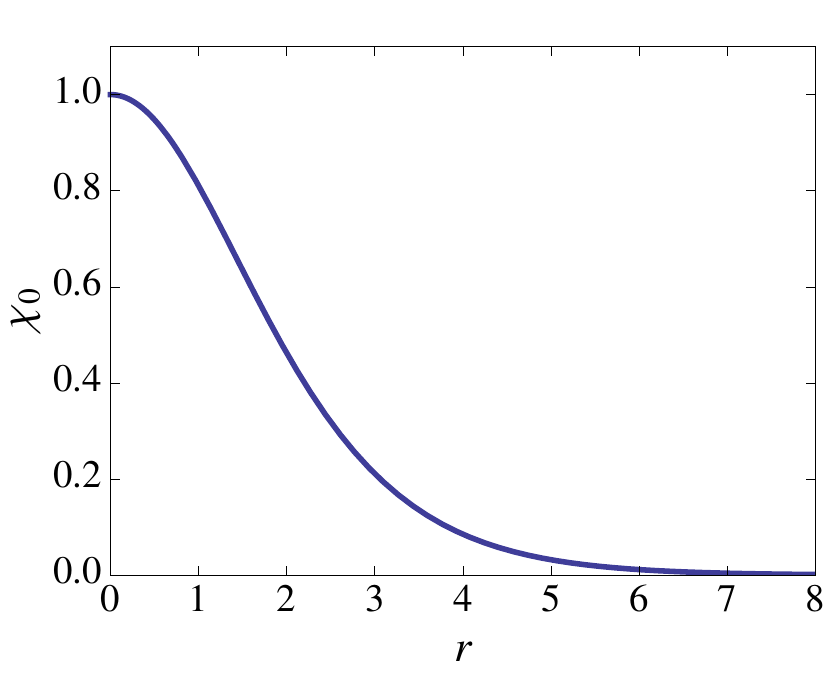}\qquad
 \includegraphics[width=0.46\textwidth]{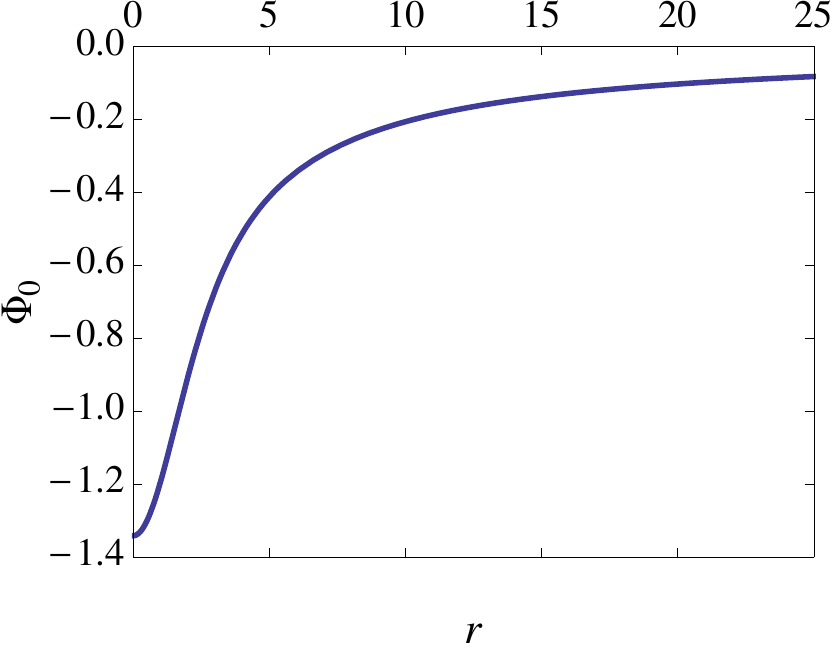}
\caption{The standard soliton profile $\chi_0(r)$ and its
  gravitational potential $\Phi_0(r)$ in dimensionless
  units.  
}
\label{fig:standsol}
\end{center}
\end{figure}

We assume spherical symmetry for soliton solutions, then the Ansatz
for a soliton
with the binding energy ${\cal E}_s$ (${\cal E}_s < 0$) takes the form  
\begin{equation} 
   \psi_s(t, {\bf x} ) = \chi (r) e^{- i{\cal E}_s t}\; ,\qquad
\Phi_s(t,{\bf x})=\Phi_\chi(r)\;,\qquad\qquad r\equiv |{\bf x}|\;.
\end{equation} 
The SP equations reduce to 
time-independent ordinary differential equations for $\chi$ and
$\Phi_\chi$. 
We impose the regular boundary
conditions of $\chi^\prime|_{r = 0} = 0 $ and $\Phi_\chi^\prime|_{r = 0} =0$  
at the center, apply the scaling symmetry of the equations to
normalize $\chi|_{r = 0}= 1$, and look for solutions with the
asymptotics $\chi|_{r  \to  \infty } = \Phi_\chi|_{r  \to  \infty }= 0$.  
The numerical solution with the number of
nodes equal to 0, as appropriate for 
the ground state, provides us with a ``standard soliton'', whose 
profile and gravitational potential are shown in  
\cref{fig:standsol}. 

The standard soliton can be rescaled to other soliton solutions through 
the scaling symmetry (\ref{eq:scaling}). Restoring the dimensionful
variables, a general soliton reads,
\be
\label{eq:gensol}
\chi=\frac{k_s^2}{\sqrt{4\pi Gm^3}}\chi_0(k_s r)~,~~~~~~
\Phi_\chi=\frac{k_s^2}{m^2}\Phi_0(k_s r)\;,
\ee 
where $k_s^{-1}$ sets its size.
The scaling of coordinates and wavefunctions leads to the scaling of
other quantities: 
the density $\rho_s \propto {\cal S}^4$, the total soliton mass $M_s =
\int d^3 {\bf x}\, \rho_s \propto {\cal S}$, 
and the binding energy $\E_s \propto {{\cal S}^2}$.
In more detail, we have
\be
\label{eq:MEs}
M_s=\mu_0 \frac{k_s}{4\pi Gm^2}~,~~~~~~
\E_s=\ve_0\frac{k_s^2}{m}\;,
\ee
where $\mu_0=25.9$ and $\ve_0=-0.692$ are the dimensionless mass and
binding energy of the standard soliton.

Let us now treat the soliton as background and 
consider small perturbations on top of it. It is convenient to factor
out the soliton time dependence and write the axion field as 
\begin{equation} 
   \psi ( t, {\bf x} ) =  \big( \chi(r) + {\delta \psi} ( t, {\bf x})
   \big)  e^{-i\E_s t} \ .   
\label{eq:psi}
\end{equation} 
We obtain the SP equations for the perturbation
by keeping only linear terms in $\delta \psi (t,{\bf x})$. Integrating
out the perturbation of the gravitational potential yields a
non-local equation,
\begin{equation}
   i \partial_t \delta \psi + \E_s  \,  \delta \psi + \frac1{2 m}
   \Delta \delta \psi-  
      m \Phi_\chi \, \delta \psi - { 4 \pi G} m^2 \chi \frac{1}{\Delta} 
         \left[ \chi \left( \delta \psi +  \delta \psi^* \right)
         \right]  = 0\;. 
   \label{eq:varphi}
\end{equation}
\Cref{eq:varphi} can be conveniently written in the following matrix form,  
\begin{equation}
 i \partial_t  
\left( \begin{array}{c} \delta \psi \\  -\delta \psi^* \end{array} \right) 
= 
\begin{pmatrix} H_0 +V  & V \\ V & H_0 +V   \end{pmatrix} 
\left( \begin{array}{c} \delta \psi \\  \delta \psi^* \end{array} \right) 
\label{eq:pertEqMatrix}
\end{equation}
and 
\begin{equation}
   H_0 = 
   - \frac \Delta  {2 m}  + (m \Phi_\chi - \E_s ) \ , 
   \quad \quad
       V =  4 \pi G m^2 \chi \frac{1}{\Delta} \chi  \ .
\label{eq:pertH0V}
\end{equation}
Note that the soliton background satisfies the equation
\be
\label{eq:solitonH0}
H_0 \,\chi = 0\;.
\ee 
From the above equations, it is clear that $\delta \psi $ and $\delta
\psi^*$ are coupled at linear order via gravity. 
As a consequence, $\delta\psi$ and $\delta\psi^*$ will contain both
positive and negative frequency components. The situation is similar
to the well-known mode mixing in atomic Bose--Einstein
condensates which implies a need for a non-trivial Bogolyubov transformation
to 
diagonalize the Hamiltonian \cite{pitaevskii2016bose}. 

If we neglect $V$ in \cref{eq:pertEqMatrix}, it reduces to
the Schr\"{o}dinger equation
$ i \partial_t  \delta \psi^{({\rm Sch})}  = H_0 \delta \psi^{({\rm
    Sch})} $, or explicitly:
\begin{equation}
   i \partial_t \delta \psi^{({\rm Sch})} + \E_s  \,  \delta \psi^{({\rm Sch})} + 
   \frac1 {2 m} \Delta \delta \psi^{({\rm Sch})}  - 
       m \Phi_\chi (r) \, \delta \psi^{({\rm Sch})}   = 0  \, .
   \label{eq:varphi2}
\end{equation}
This does not have mixing between positive and negative frequency
modes.
Neglecting $V$ is in general not justified since inside the soliton
$V$ and $m\Phi_\chi$ are of the same order. The low-lying bound states
whose wavefunctions significantly overlap with the soliton will be
affected comparably by these two terms. On the other hand, one expects
the Schr\"odinger approximation to become more accurate for highly
excited states with wavefunctions wider than the soliton. Indeed, the
Newtonian potential has a long tail, $\Phi_\chi=-GM_s/r$ outside the
soliton, whereas the operator $V$ is proportional to the soliton profile
$\chi$ and hence is 
sharply localized at distances
$r\lesssim k_s^{-1}$. Therefore, $V$ is expected to
have weak effect on 
the dynamics of high-level modes. Our numerical results below will
confirm these expectations.

\section{General Properties of Eigenstates}
\label{sec:perturbations}

In this section we study the coupled positive and negative frequency modes
governed by the linearized SP equations and discuss
their general properties,  
including the orthogonality and completeness relations.

\subsection{Positive, negative, and zero frequency modes}

\Cref{eq:pertEqMatrix,eq:pertH0V} show that the positive and negative
frequency modes are coupled. 
This coupling guides us to 
look for eigenmodes of the form,
\begin{equation}
\label{eq:modeform}
  {\delta \psi }_a  ( t, {\bf x} ) =  
      \varphi_a^{(1)} ({\bf x} ) \, e^{- i \epsilon_a t }  
      + {\varphi_a^{(2)}}^* ( {\bf x} ) \, e^{ +  i \epsilon_a
        t}\;,~~~~~~
\epsilon_a\geq 0  \ .
\end{equation}
Here the subscript $a$ labels different energy levels and
$\epsilon_a$ is the energy of the excitation relative to the soliton ground
state. 
The excitations are bound to the soliton as long as their energy is
below the soliton binding energy,
$\epsilon_a<|\E_s|$; otherwise they describe a continuum spectrum of
axion waves propagating to infinity.

Due to
the spherical symmetry of the soliton background, 
$a$ is a set of orbital, radial, and magnetic quantum number,
$a=( \ell, n, m )$,  
as we factorize $\varphi$ into the radial and angular components,
\begin{equation}
\varphi_{ \ell n m} ( r , \theta, \phi) = \varphi_{ \ell n} (r)\,  
Y_{\ell m}( \theta, \phi)\;.
\label{eq:decompvarphi}
\end{equation}
Here $Y_{\ell m}$ are the spherical
harmonics.\footnote{Following the standard convention, we denote the
  magnetic quantum number by $m$, same as the axion mass. This should
  not lead to confusion since the magnetic
  quantum number almost never 
explicitly appears in the formulas.} 
The non-negative integers $n\geq 0$ label 
the discrete energy levels at a given 
orbital quantum number $\ell$.\footnote{
The radial quantum number $n$ should not be confused with the principal
quantum number $n_p$ of atomic physics. For the hydrogen atom the two
are related as $n_p=\ell+n+1$ and the energy levels depend only on
this combination, $E_{\ell n}=-\frac{1}{2(\ell+n+1)^2}$ (in atomic
units). In the soliton background the dependence of the energy levels
on $\ell$ and $n$ is more complicated and can be found only numerically.
Nevertheless, we will see that some approximate hydrogen-like
degeneracies arise between the energies of highly-excited states.} 
For the continuum spectrum, we can choose instead of $n$ the absolute
value of the mode momentum at infinity.  
The SP equations (\ref{eq:pertEqMatrix}) are
converted to coupled equations  
of $\varphi_a^{(1)}$ and $\varphi_a^{(2)}$, 
\begin{equation}
\begin{pmatrix} H_0 +V  & V \\ V & H_0  +V \end{pmatrix} 
\left( \begin{array}{c} \varphi_a^{(1)} \\  \varphi_a^{(2)}  \end{array} \right) 
= 
\epsilon_a 
\left( \begin{array}{c} \varphi_a^{(1)} \\  - \varphi_a^{(2)}  \end{array} \right)  \ .
\label{eq:pertEqMatrix2}
\end{equation}
These can be cast in a compact form,
\begin{equation}
\sigma_3 H \, \varphi_a
= 
\epsilon_a  \, \varphi_a \ , 
\quad \text{with} \quad
H = 
\begin{pmatrix} H_0 + V & V \\ V & H_0 +V   \end{pmatrix},~~~~~ 
\varphi_a = 
\left( \begin{array}{c} \varphi_a^{(1)} \\
    \varphi_a^{(2)}  \end{array} \right),  
\label{eq:pertEqMatrixComp}
\end{equation}
and $\sigma_3 = diag ( 1, -1)$ being the third Pauli matrix. In this
way we have reformulated the problem of finding the spectrum of
excitations in the presence of the soliton as an eigenvalue problem in
the space of two-component columns $\vf$, with $\vf^{(1)}$
corresponding to positive and $\vf^{(2)}$ to negative
frequencies. 

Note, however, that the effective Hamiltonian $\sigma_3
H$ is neither Hermitian, nor positive definite. Instead, its spectrum
is symmetric around zero. To see this observe that $H$ commutes with
the first Pauli matrix, $[\sigma_1,H]=0$. Then, given a state $\vf_a$
with positive energy, we can construct its dual
\be
\label{eq:vfdual}
\tilde\vf_a=\sigma_1\vf_a^*\;,~~~~~\epsilon_a>0\;,
\ee
with the negative eigenvalue,
$\sigma_3H\tilde\vf_a=-\epsilon_a\tilde\vf_a$. Clearly, the duality
interchanges the positive and negative frequency components. Comparing
with \cref{eq:modeform}, we see that the dual state corresponds to the
complex conjugate axion wavefunction.

The Hamiltonian also has zero-energy eigenstates which 
deserve a separate discussion. These are the
Goldstone modes associated to the symmetries of the axion dynamics
spontaneously broken by the soliton. It is straightforward to identify
these symmetries from the axion action
\begin{equation}
   S = \int d^4 x  \,\bigg( i \psi^* \partial_t \psi 
+ \frac{1}{2 m}  \psi^* \Delta \psi  
      - 2 \pi G m |\psi|^2 \frac{1}{ \Delta} |\psi|^2\bigg)\;,
\end{equation}
where we have integrated out the gravitational potential $\Phi$. It is
invariant under the $U(1)$ phase shifts of the complex scalar field,
as well as under spatial translations. Acting with these symmetries on
the soliton we obtain other solutions of the field equations with the
same energy,  
\begin{equation}
\psi_{s,\alpha}(t, {\bf x}) = \chi( {\bf x}) \, e^{i \alpha} \, e^{- i \E_s t}   \, , 
\quad \quad 
\psi_{s,{\bf a}}(t, {\bf x}) = \chi( {\bf x} + {\bf a}) \,  e^{- i \E_s t}    \, .
\end{equation}
The Goldstone modes are the infinitesimal versions of these
transformations giving zero-energy states in the monopole and dipole
sectors, which in the two-component form read,  
\begin{equation} 
   { \varphi}_{00} ( {\bf x} ) = 
\left( \begin{array}{c} \chi ( r)  \\  - \chi (r)   \end{array} \right)  \ , 
   \quad \quad
   { \varphi}_{10} ( {\bf x} ) =  
\left( \begin{array}{c} \nabla \chi ( r)  \\  \nabla \chi (r)   \end{array} \right)  \ .
\label{eq:groundstate}
\end{equation} 
Recall that the first number in the subscript stands for the angular
momentum $\ell=0$ or $\ell=1$, and the second number $0$ stands for
the ground state in the corresponding sector. Note that, with a slight
abuse of notations, we have denoted by  
${\varphi}_{10} ( {\bf x} )$ a triplet of modes differing by the
magnetic quantum number $m=-1,0,1$; this triplet can also be
written as $
\left(  \partial_r \chi ( r)  ,  \partial_r \chi (r)    \right)^T  \,
Y_{1m}( \theta, \phi )$. 

It is instructive to check that the modes (\ref{eq:groundstate})
indeed satisfy \cref{eq:pertEqMatrixComp} with zero r.h.s. by a direct
substitution. For the monopole $\vf_{00}$ the terms with $V$ cancel
and we arrive at the background equation (\ref{eq:solitonH0}). For the
dipole, we act with the derivative $\nabla$ on \cref{eq:solitonH0} and
obtain $( H_0  + 2V ) \nabla \chi = 0  $, which is precisely the
equation given by substituting $\vf_{10}$ into
\cref{eq:pertEqMatrixComp}. This reasoning also tells us that the
Schr\"odinger equation (\ref{eq:varphi2}) has 
the monopole zero mode, but not the dipole zero mode, 
because it neglects $V$ and $ H_0\nabla \chi \neq 0$.

\subsection{Orthogonality and completeness relations}

Since the operator in the eigenvalue problem (\ref{eq:pertEqMatrixComp}) is
not Hermitian, it leads to unusual orthogonality and completeness
relations. 
By using \cref{eq:pertEqMatrixComp} and the identity 
$ \int d^3{\bf x} \big( ( \varphi_b^\dagger H \varphi_a ) - (
  \varphi_a^\dagger H \varphi_b )^\dagger  \big) = 0 $,  
we find the first orthogonality condition 
\begin{equation}
   ( \epsilon_a - \epsilon_b ) \, \int d^3 {\bf x} \,
   \varphi_b^\dagger ( {\bf x} ) \, \sigma_3 \, \varphi_a ({\bf x}) =
   0   \ . 
\label{eq:orthcond1}
\end{equation}
Assuming $\epsilon_a \neq \epsilon_b$ we can write the condition
in terms of  
positive and negative frequency components, 
\begin{equation}
\int d^3 {\bf x} \, 
\left[ {\varphi_b^{(1)}}^* ( {\bf x} )  \, \varphi_a^{(1)} ( {\bf x} )  
 - {\varphi_b^{(2)}}^* ( {\bf x} )  \, \varphi_a^{(2)} ( {\bf x} )
\right]   = 0 \ , \quad  \epsilon_a \neq \epsilon_b \ . 
\label{eq:orthcond1_v2}
\end{equation}
Note the minus sign in front of the negative-frequency contribution. 
Likewise, we use \cref{eq:pertEqMatrixComp} and another identity   
$ \int d^3{\bf x} \left( ( \varphi_b^T \sigma_1  H \varphi_a )  - (
  \varphi_a^T \sigma_1 H \varphi_b )^T  \right) = 0 $  
to obtain the second orthogonality condition,
\begin{equation}
   ( \epsilon_a  +  \epsilon_b ) \,  \int d^3{\bf x} \,  \varphi_b^T  ( {\bf x} ) \, \sigma_1 \sigma_3 \, \varphi_a ({\bf x}) 
       = 0     \ . 
\label{eq:orthcond2}
\end{equation}
In terms of components, it reads
\begin{equation}
\int d^3{\bf x} \,  \left[ 
{\varphi_b^{(1)}} ( {\bf x} )  \, \varphi_a^{(2)} ( {\bf x} )  
 - {\varphi_b^{(2)}} ( {\bf x} )  \, \varphi_a^{(1)} ( {\bf x} ) \right]   = 0 \ , 
\quad  \epsilon_a +\epsilon_b \neq 0  \, .
\label{eq:orthophi12c}
\end{equation}
Note that it is also satisfied for the zero modes
(\ref{eq:groundstate}), despite the fact that for them
$\epsilon_a=\epsilon_b=0$. 

The first orthogonality condition (\ref{eq:orthcond1})  
implies that we can set the inner product and  
normalization of positive-energy wavefunctions by using $\sigma_3$,  
\begin{equation}
\langle \vf_a,\,\vf_b\rangle\equiv   
\int d^3 {\bf x} \, \varphi_a^\dagger ( {\bf x} ) \, \sigma_3 \,
   \varphi_b( {\bf x}) = \delta_{ab}\;,~~~~~~\epsilon_a>0  \ . 
\label{eq:basisNorm1}
\end{equation}
Their dual states (\ref{eq:vfdual}) corresponding to negative
eigenvalues $-\epsilon_a$ then have a negative norm and are orthogonal
to all $\vf_b$, 
\begin{equation}
   \langle \tilde\varphi_a , \,  \tilde \varphi_b\rangle = - \delta_{ab}\;,
\qquad\qquad \langle \tilde\varphi_a,\, \varphi_b\rangle = 0
          \ .
   \label{eq:basisNorm2}
\end{equation}

The normalization condition (\ref{eq:basisNorm1}) cannot be applied to
the ground states (\ref{eq:groundstate}) since their norm vanishes,
\begin{equation}
 \langle \varphi_{00}, \, \varphi_{00}\rangle 
= \langle \varphi_{10},\, \varphi_{10} \rangle   = 0 \, .
   \label{eq:0product}
\end{equation}
This raises an issue of the completeness of the eigenmode
basis. Indeed, consider e.g. a column $\vf=(\chi(r),0)^T$. If it were
possible to decompose it as a linear combination of eigenmodes, 
\be
\label{eq:vflinear}
\varphi = \sum_{\epsilon_a\geq 0} \alpha_a \varphi_a +
\sum_{\epsilon_a> 0} \tilde \alpha_a \tilde \varphi_a \;,
\ee
its product with $\vf_{00}$ would be zero. However, the explicit
evaluation gives
\be
\langle\vf,\,\vf_{00}\rangle=\int d^3{\bf x}\,\chi^2(r)>0\;.
\ee
Thus the expansion (\ref{eq:vflinear}) must be supplemented by a basis
state having non-zero inner product with $\vf_{00}$. We will refer to
this vector as conjugate to $\vf_{00}$ and will denote it by
$\hat\vf_{00}$. Repeating the argument for  $\vf=(\nabla\chi(r),0)^T$ we conclude that we also need to include a vector
$\hat\vf_{10}$ conjugate to $\vf_{10}$.

To find $\hat\vf_{00}$, $\hat\vf_{10}$ we again use the symmetries of
the problem. Recall that the SP system is
invariant under the scaling (\ref{eq:scaling}) which is broken by the
soliton background. The scaling transformation changes the binding
energy of the soliton and hence does not give rise to a Goldstone
mode. Instead, it provides us with $\hat\vf_{00}$. To see this,
apply an infinitesimal transformation ${\cal S}=1+\alpha$, $\alpha\ll
1$, to the soliton. This generates a perturbation,  
\begin{equation}
\label{eq:scalemode}
   \delta\psi (t, {\bf x}) =  \alpha \,\big(2 \chi (r) + 
{\bf x} \cdot \nabla \, \chi (r) 
-  i 2 \E_s  \, \chi(r)\,t \big) \ .
\end{equation}
Note that it linearly grows with time. Since both the rescaled and the
original solitons satisfy the SP system, the
perturbation (\ref{eq:scalemode}) obeys the dynamical equation
(\ref{eq:pertEqMatrix}). Substituting and using $H\vf_{00}=0$ we
obtain,
\begin{equation}
\label{eq:vf00hat}
 \sigma_3  H \, \hat{\varphi}_{00}   ({\bf x})= \varphi_{00}  ({\bf x})  \, , 
   \quad \text{where}\quad
   \hat{\varphi}_{00}  ({\bf x})
 = \frac{1}{2\E_s} 
\left( \begin{array}{c}  2\chi (r) + {\bf x} \cdot \nabla    \, \chi (r) \\  
   2\chi (r) + {\bf x} \cdot \nabla    \, \chi (r)
 \end{array} \right) 
\end{equation}
From $(\sigma_3  H)^2\,  \hat{\varphi}_{00} = \sigma_3H \,
{\varphi}_{00}   = 0 $ 
and 
$(\sigma_3  H)^2 \,  \varphi_{a} = \epsilon_a^2 \,\varphi_a$, 
we infer that $\hat{\varphi}_{00}$ is 
orthogonal to all eigenmodes with strictly positive energies and their duals. 
Clearly, its inner product with itself also vanishes. On the other
hand, $\hat\vf_{00}$ has non-zero inner product with $\vf_{00}$,   
\begin{equation}
   \langle \hat {\varphi}_{00},\, \varphi_{00}\rangle
         = \frac{1}{\E_s}\int  d^3{\bf x}  \, \left(  2 \chi^2(r)  
 + \frac{1}{2} \, {\bf x} \cdot \nabla \, \chi^2(r) \right)  
      =  \frac{1}{2 \E_s}  \int d^3{\bf x} \, \chi^2 (r) \equiv  C_{00}\;,
\end{equation}
and thus is the sought-after conjugate state to $\vf_{00}$. Note that
\cref{eq:vf00hat} implies that in the subspace spanned by $\vf_{00}$
and $\hat\vf_{00}$ the effective Hamiltonian $\sigma_3 H$ has the form
of a Jordan cell $\left(\begin{smallmatrix}0&1\\0&0\end{smallmatrix}\right)$. To
summarize, the orthogonality relations for the sates in the   
$(00)$ sector are
\begin{align}
& \langle   \varphi_{00},\,\varphi_{00} \rangle
= \langle\hat\varphi_{00},\,\hat\varphi_{00}\rangle = 0 \, , 
   \quad 
    \langle\hat\varphi_{00},\, \varphi_{00}\rangle = C_{00}  \, , 
   \nonumber
   \\
 &  \langle\varphi_{00},\, \varphi_{a}\rangle 
=\langle \hat{\varphi}_{00},\,\varphi_{a} \rangle = 0 \, 
   \quad
   {\rm for} \, \,  \epsilon_a \neq 0 \,  . 
\end{align}

One more symmetry spontaneously broken by the soliton is the Galilean
invariance, under which the axion field changes as
\be
\label{eq:Galilean}
\psi(t,{\bf x})\to e^{-im\,{\bf v}\cdot {\bf x}-i\frac{m v^2}{2}t}
\,\psi(t,{\bf x}+{\bf v}t)\;.
\ee
Applying it to the soliton, we find a linearly growing perturbation,
\begin{equation}  
     \delta\psi ( t, {\bf x} ) = {\bf v}\cdot 
\big(- i m\, {\bf x}\, \chi(r) + \nabla \chi (r)\,t\big)  \, .
\end{equation}   
Substituting into \cref{eq:pertEqMatrix} we arrive at  
\begin{equation}
\label{eq:hatvf10}
\sigma_3   H \, \hat{\varphi}_{10} = \varphi_{10}  \;\;\; 
   \quad\text{with} \quad
   \hat{\varphi}_{10} 
 = m\left( \begin{array}{c} -{\bf x}  \, \chi (r) \\  {\bf x} \,  \chi
     (r) \end{array} \right)  \ .  
\end{equation}
We see that in the $(10)$ sector the effective Hamiltonian $\sigma_3H$
is a also Jordan cell. Reasoning as before, we conclude that the state
$\hat\vf_{10}$ has zero norm and is orthogonality to all eigenstates of
$\sigma_3H$, except $\vf_{10}$ for which we have,
\begin{equation}
  \langle\hat{\varphi}_{10i},\,\varphi_{10j} \rangle
         = -m\int  d^3{\bf x}  \, x_i \nabla_j \chi^2(r)  = 
m\delta_{ij} \int d^3{\bf x} \, \chi^2 (r) \equiv
\delta_{ij}C\,_{10}~,~~~~~
i,j=1,2,3\;, 
\end{equation}
where we have recalled that $\vf_{10}$ and $\hat\vf_{10}$ are actually
triplets of states.
As a summary, the orthogonal conditions for the $(10)$ sector are 
\begin{align}
&      \langle    \varphi_{10i},\, \varphi_{10j}\rangle=
\langle\hat\varphi_{10i},\,\hat\varphi_{10j}\rangle= 0 \, , 
   \quad 
    \langle\hat\varphi_{10i},\, \varphi_{10j}\rangle =\delta_{ij}\, C_{10}  \, , 
   \nonumber
   \\
&   \langle \varphi_{10i},\,\varphi_{a}\rangle
=\langle\hat{\varphi}_{10i},\, \varphi_{a}\rangle = 0 \, 
   \quad
   {\rm for} \, \,  \epsilon_a \neq 0 \,  . 
\label{eq:10orghogoal}
\end{align}

We have now constructed the complete basis in the space of
two-component columns. It consists of the eigenmodes of $\sigma_3H$
and the two conjugate states $\hat\vf_{00}$ and $\hat\vf_{10}$. An
arbitrary column can be represented as 
\begin{equation}
\label{eq:decomptrue}
      \varphi ( {\bf x} )  =  
 \sum_{\epsilon_a > 0 }\big(  \alpha_a \, \varphi_a({\bf x})
 + \tilde \alpha_a \, \tilde\varphi_a({\bf x})\big)
       + \alpha_{00}\, \varphi_{00}({\bf x}) 
       + \hat\alpha_{00}\, \hat\varphi_{00} ({\bf x})
       + \sum_{i}\big(\alpha_{10i}\, \varphi_{10i}({\bf x})
       + \hat\alpha_{10i}\, \hat\varphi_{10i}({\bf x})\big)\;,
\end{equation}
and the coefficients in this expansion are obtained by taking the
inner product of $\vf$ with the basis states:
\begin{gather}
      \alpha_a = \langle\varphi_a, \,  \varphi\rangle   \, ,
      \quad 
      \tilde\alpha_a = - \langle \tilde\varphi_a,  \, \varphi\rangle  \, ,\notag\\
      \alpha_{00} = \frac{\langle\hat\varphi_{00},\,\varphi\rangle}{C_{00}}  \, , 
      \quad
      \hat\alpha_{00} =
      \frac{\langle\varphi_{00},\,\varphi\rangle}{C_{00}} \, ,
\quad
      \alpha_{10} = \frac{\langle\hat\varphi_{10},\, \varphi\rangle}{C_{10}}  \, , 
      \quad
      \hat\alpha_{10} = \frac{\langle \varphi_{10},  \, \varphi\rangle}{C_{10}}  \, .
\end{gather}
This implies the following decomposition of the $\delta$-function:
\begin{equation}
\label{eq:unitfull}
   \begin{split}
\sigma_3\, \delta({\bf x } - {\bf y} ) = &
\sum_{\epsilon_a >0} \big(\varphi_a ({\bf x}) \otimes \varphi_a^\dagger({\bf y})
      - \tilde\varphi_a({\bf x}) \otimes\tilde \varphi_a^\dagger ({\bf
        y})\big)
      \\
   &+ \frac{1}{C_{00}} \big(\varphi_{00} ({\bf x}) 
\otimes \hat\varphi_{00}^\dagger ({\bf y})  
   + \hat \varphi_{00} ({\bf x}) \otimes \varphi_{00}^\dagger ({\bf
     y}) \big)\\ 
    &  + \frac{1}{C_{10}} \sum_i\big(\varphi_{10i} ({\bf x}) \otimes 
\hat\varphi_{10i}^\dagger ({\bf y})  
   +  \hat\varphi_{10i} ({\bf x}) \otimes \varphi_{10i}^\dagger ({\bf
     y}) \big) \;.
   \end{split}
\end{equation}

Returning to the physical axion perturbations $\delta\psi$, we observe
from \cref{eq:pertEqMatrix} that they correspond to self-dual
columns. This imposes restrictions on the coefficients in
\cref{eq:decomptrue}: $\tilde\alpha_a$ must be complex conjugate of
$\alpha_a$, whereas $\alpha_{00}$, $\hat\alpha_{10}$ must be purely
imaginary and $\hat\alpha_{00}$, $\alpha_{10}$ real. Restoring the
time dependence, we obtain the most general axion perturbation in the
soliton background, 
\be
\label{eq:gensolut}
\begin{split}
\delta\psi(t,{\bf x})=&
\sum_{\epsilon_a > 0 }\Big(\alpha_a \, \varphi^{(1)}_a({\bf
  x})e^{-i\epsilon_a t}
 + \alpha^*_a \, {\varphi^{(2)}_a}^*({\bf x})
e^{i\epsilon_a t}
\Big)\\
   &    + \alpha_{00}\, \chi(r) 
       + \hat\alpha_{00}\bigg[\frac{1}{2\E_s}
\big(2\chi(r)+{\bf x}\cdot\nabla\chi(r)\big)-i\chi(r)\,t\bigg]\\
 &   + \alpha_{10i}\, \nabla_i\chi(r)
       + \hat\alpha_{10i}\big[-m\,{\bf x}_i\,\chi(r)
-i\nabla_i\chi(r)\,t\big]\;,
\end{split}
\ee
with $\Re{\alpha_{00}}=\Re{\hat\alpha_{10i}}=0$, 
$\Im{\hat\alpha_{00}}=\Im{\alpha_{10i}}=0$.

Finally, let us note that for the eigenmodes of the Schr\"odinger
equation (\ref{eq:varphi2}) the orthogonality and completeness
relations are standard, 
\begin{equation}
\label{eq:Schrnorm}
   \int d^3 {\bf x} \, {\varphi_a^{({\rm Sch})}}^*  ( {\bf x}) \,
   \varphi^{({\rm Sch})}_b ( {\bf x} ) =  \delta_{ab} \;, \qquad\qquad 
  \sum_{a } \varphi^{\rm (Sch)}_{ a  }  ({\bf x}) \, {\varphi^{\rm (Sch)}_{ a  }}^*
      ( {\bf y})    
      =  \delta ( {\bf x}- {\bf y} )   \, ,
\end{equation}
so that any solution of \cref{eq:varphi2} is expanded as
\be
\delta\psi^{({\rm Sch})}(t,{\bf x})=\sum_a\beta_a\,\varphi_a^{({\rm
    Sch})}({\bf x})\,e^{-i\epsilon^{\rm (Sch)}_at}\;.
\ee
The energies $\epsilon^{\rm (Sch)}_a$ here are the eigenvalues
satisfying 
\be
\label{eq:Schrop}
H_0\,\varphi_a^{({\rm
    Sch})}=\epsilon^{\rm (Sch)}_a\,\varphi_a^{({\rm
    Sch})}
\ee 
and are in general different from $\epsilon_a$.

\section{Energy Eigenstates}
\label{sec:PertTheory}

In this section we present the numerical results for the energy
eigenstates in the soliton background and compare the levels and
wavefunctions of the full SP system with those of the Schr\"odinger
equation in the fixed gravitational potential. 
We work in
dimensionless units $m=4\pi G=1$ and choose the standard soliton
discussed in \cref{sec:basicEq} as the background. Solutions for any
other soliton can be obtained by the rescaling (\ref{eq:scaling}).

\subsection{Numerical results for the Schr\"{o}dinger--Poisson and
  Schr\"odinger equations}

We solve numerically the eigenvalue problem (\ref{eq:pertEqMatrix2})
in various multipole sectors labeled by the orbital number $\ell$. To
resolve the non-locality of the operator $V$ (see \cref{eq:pertH0V}),
we integrate back in the perturbation of the gravitational potential which
obeys the equation
$\Delta\delta\Phi=\chi\,(\delta\psi+\delta\psi^*)$. Decomposing it
into spherical harmonics similarly to \cref{eq:decompvarphi}, we
arrive at the system of 3 coupled second-order ODEs: 
\bseq 
\label{eq:speq}
\begin{align}
    & \varphi^{(1) \, \prime \prime}_{\ell n}(r)  
     + \frac{2}{r} \,  \varphi_{\ell n}^{(1) \, \prime}(r) 
         - 2 \left( \Phi_0 (r) -\ve_0 - \epsilon_{\ell n}  + \frac{ \ell ( \ell + 1 ) }{ 2 \, r^2} \right)  \varphi^{(1)}_{\ell n}(r)  
         =2 \, \delta\Phi_{\ell n }(r) \chi_0  (r) 
\label{eq:speq1}
      \\
   &
   \varphi^{(2) \, \prime \prime}_{\ell n}(r)  
     + \frac{2}{r} \,  \varphi_{\ell n}^{(2) \, \prime}(r) 
         - 2 \left( \Phi_0 (r) - \ve_0 + \epsilon_{\ell n}  + \frac{ \ell ( \ell + 1 ) }{ 2 \, r^2} \right)  \varphi^{(2)}_{\ell n}(r)  
         =2 \, \delta \Phi_{\ell n }(r) \chi_0  (r) 
\label{eq:speq2}
      \\
  & 
      \delta \Phi_{\ell n }''(r)
      + \frac{2}{r} \delta \Phi_{\ell n }'(r)
      -  \frac{ \ell ( \ell + 1 ) }{  r^2} \delta \Phi_{\ell n } (r) 
         = \chi_0 (r)  \left( 
          \varphi^{(1)}_{\ell n}(r) 
          + \varphi^{(2)}_{\ell n}(r) 
         \right)   
\label{eq:speq3}
\end{align}
\eseq
Here $\chi_0$, $\Phi_0$ and $\ve_0$ are the profile, gravitational
potential and the binding energy of the standard soliton. 

For the Schr\"odinger equation (\ref{eq:Schrop}) we obtain a single
ODE,
\begin{equation}
\varphi_{\ell n}^{\prime \prime \rm(Sch)}(r) + \frac{2}{r} \,
\varphi_{\ell n}^{\prime \rm(Sch)}(r)  
-2\bigg(\Phi_0 (r) -\ve_0- \epsilon_{\ell n}^{\rm (Sch)}
+\frac{ \ell ( \ell + 1 ) }{2 r^2}\bigg)\varphi^{\rm(Sch)}_{\ell n}(r)  = 0 \;.  
\label{eq:ODE}
\end{equation}
We solve the system (\ref{eq:speq}) and \cref{eq:ODE} numerically
subject to the regularity conditions at the center $r=0$ and
vanishing conditions at infinity. Under these conditions all
wavefunctions can be chosen real. The details of our numerical
procedure are described in \cref{app:H0}.

\afterpage{\clearpage}

\begin{figure}[!tp]
\begin{center}
~\vspace{-0.5cm}
\\
\includegraphics[width=0.3\columnwidth]{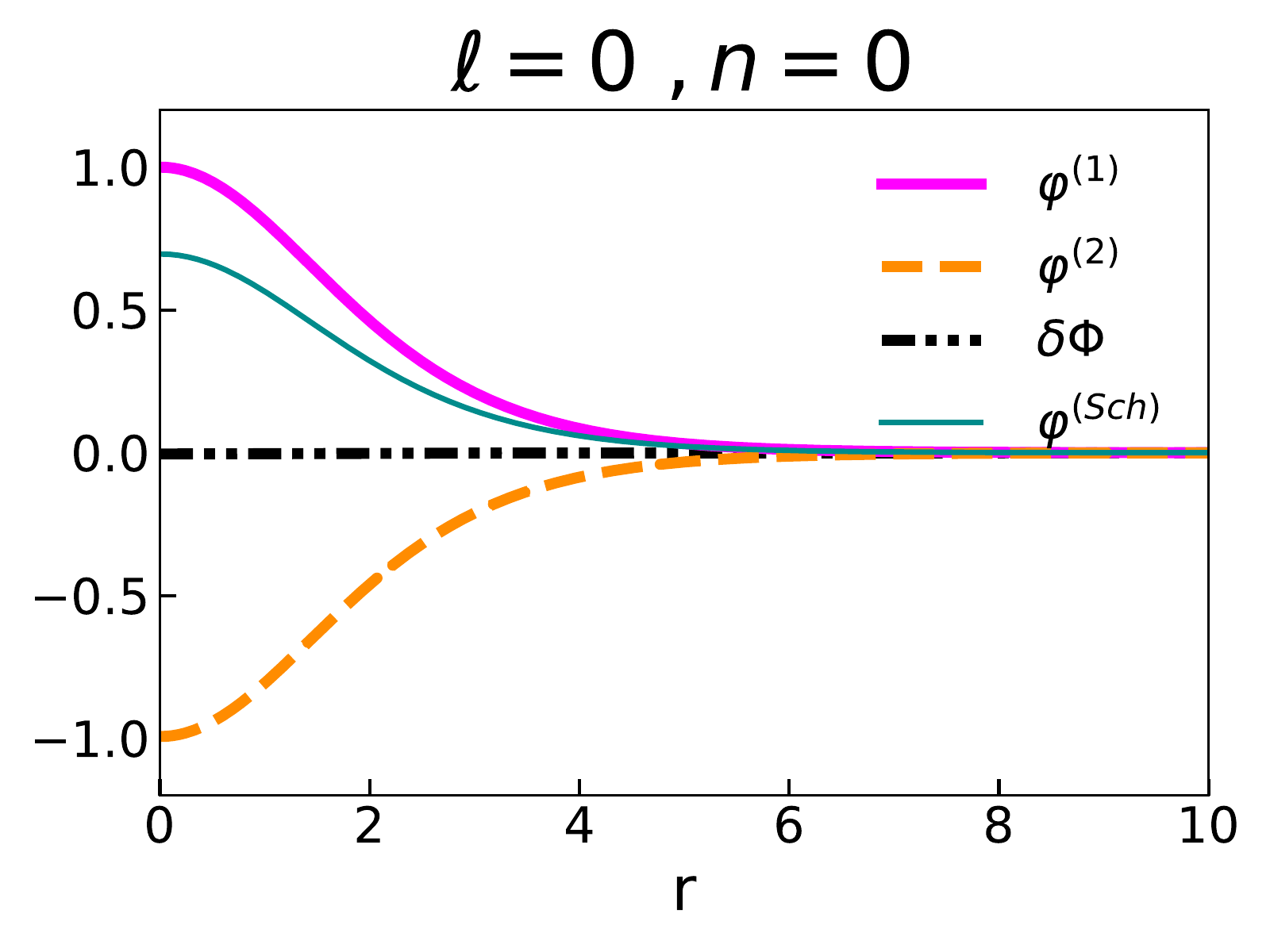}
\includegraphics[width=0.3\columnwidth]{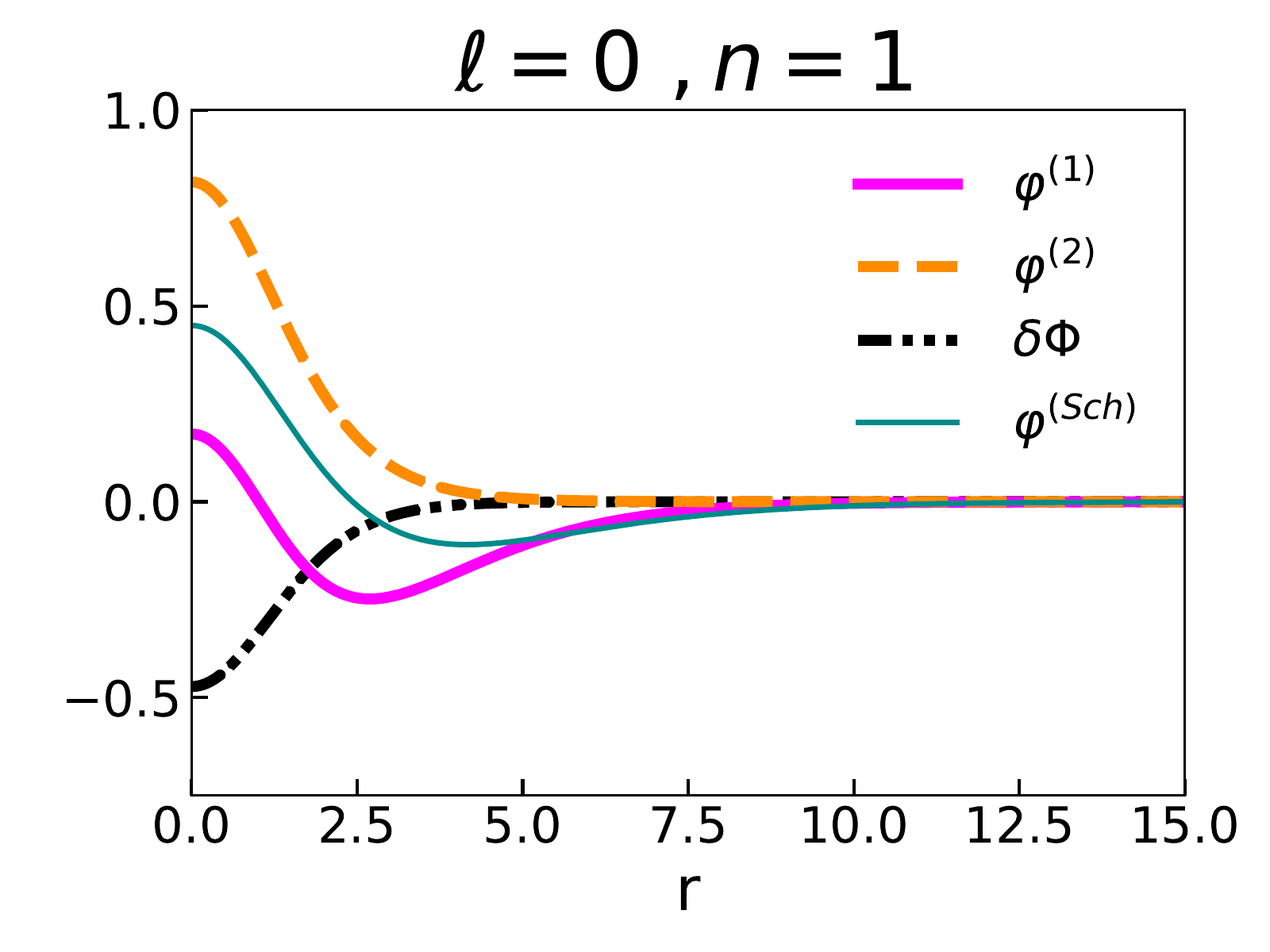}
\includegraphics[width=0.3\columnwidth]{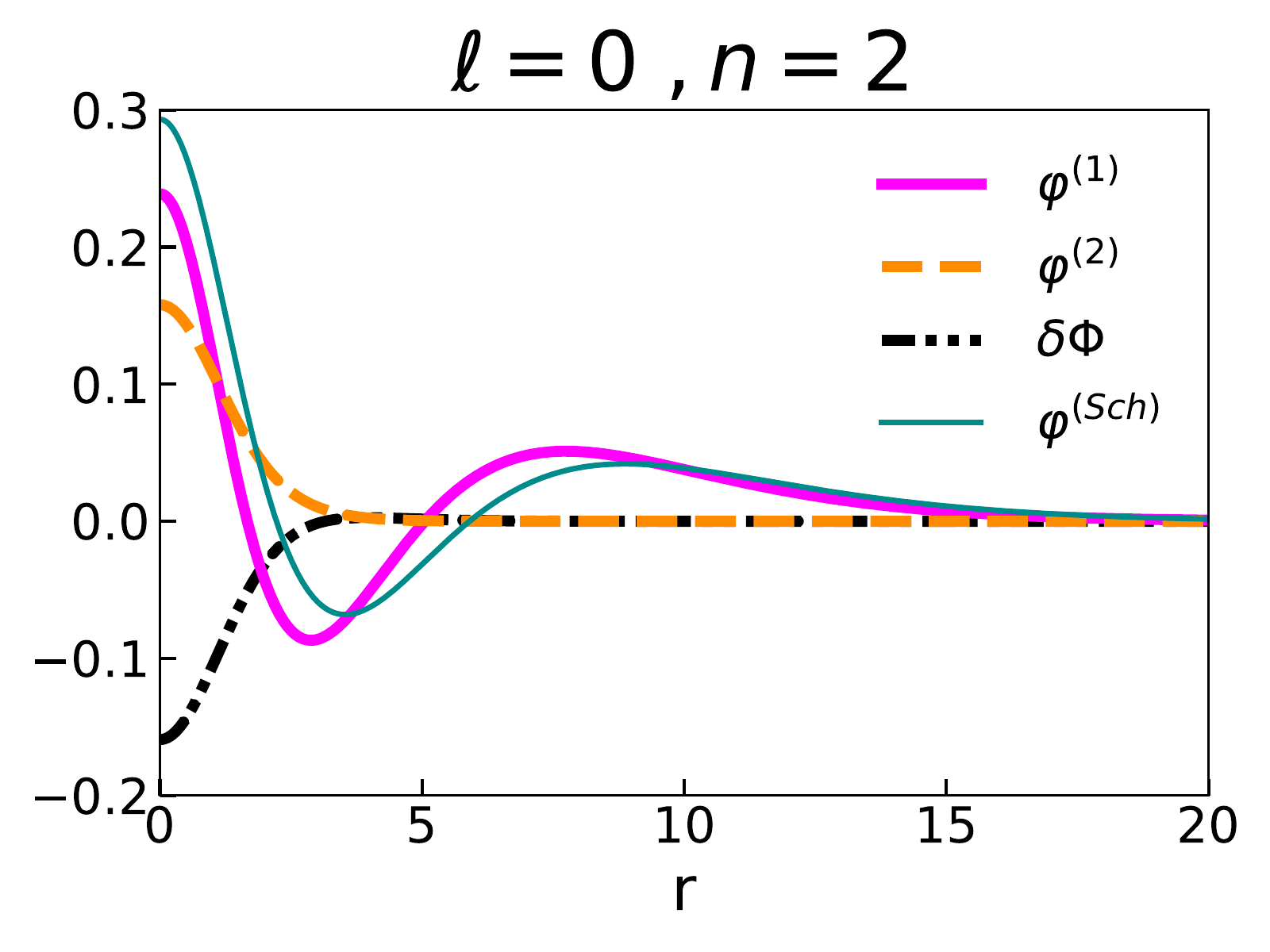}
\includegraphics[width=0.3\columnwidth]{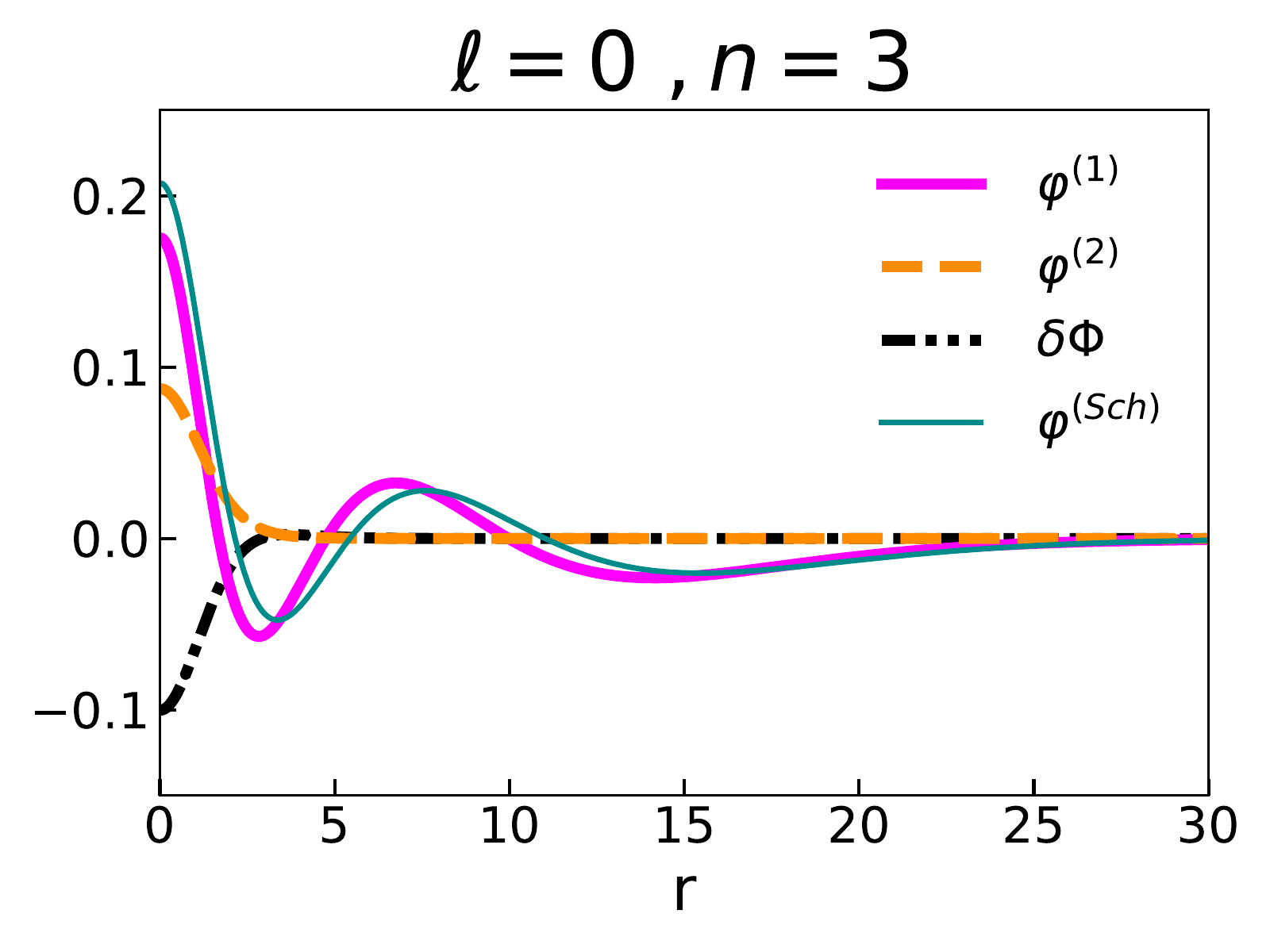}
\includegraphics[width=0.3\columnwidth]{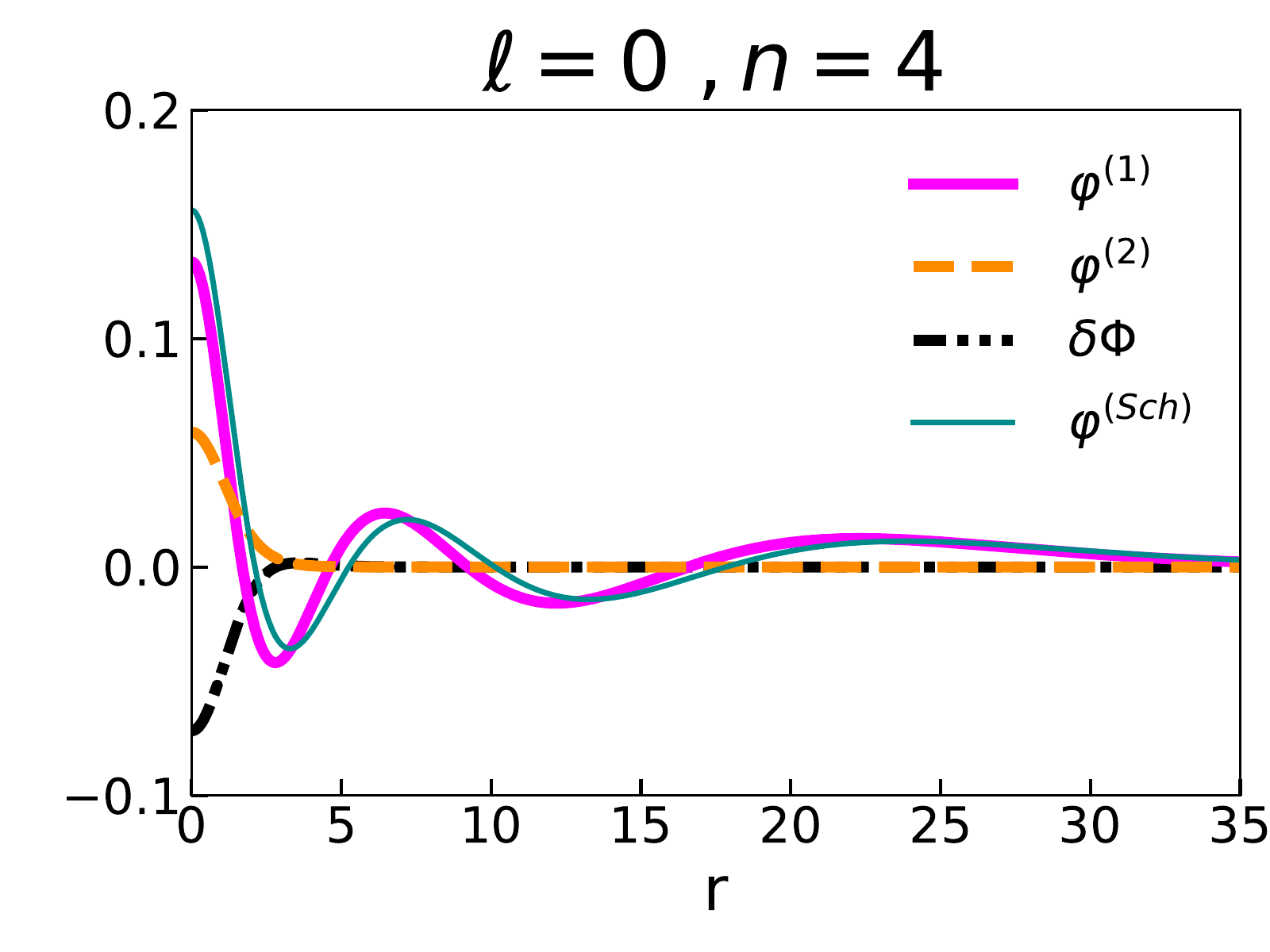}
\\
\includegraphics[width=0.3\columnwidth]{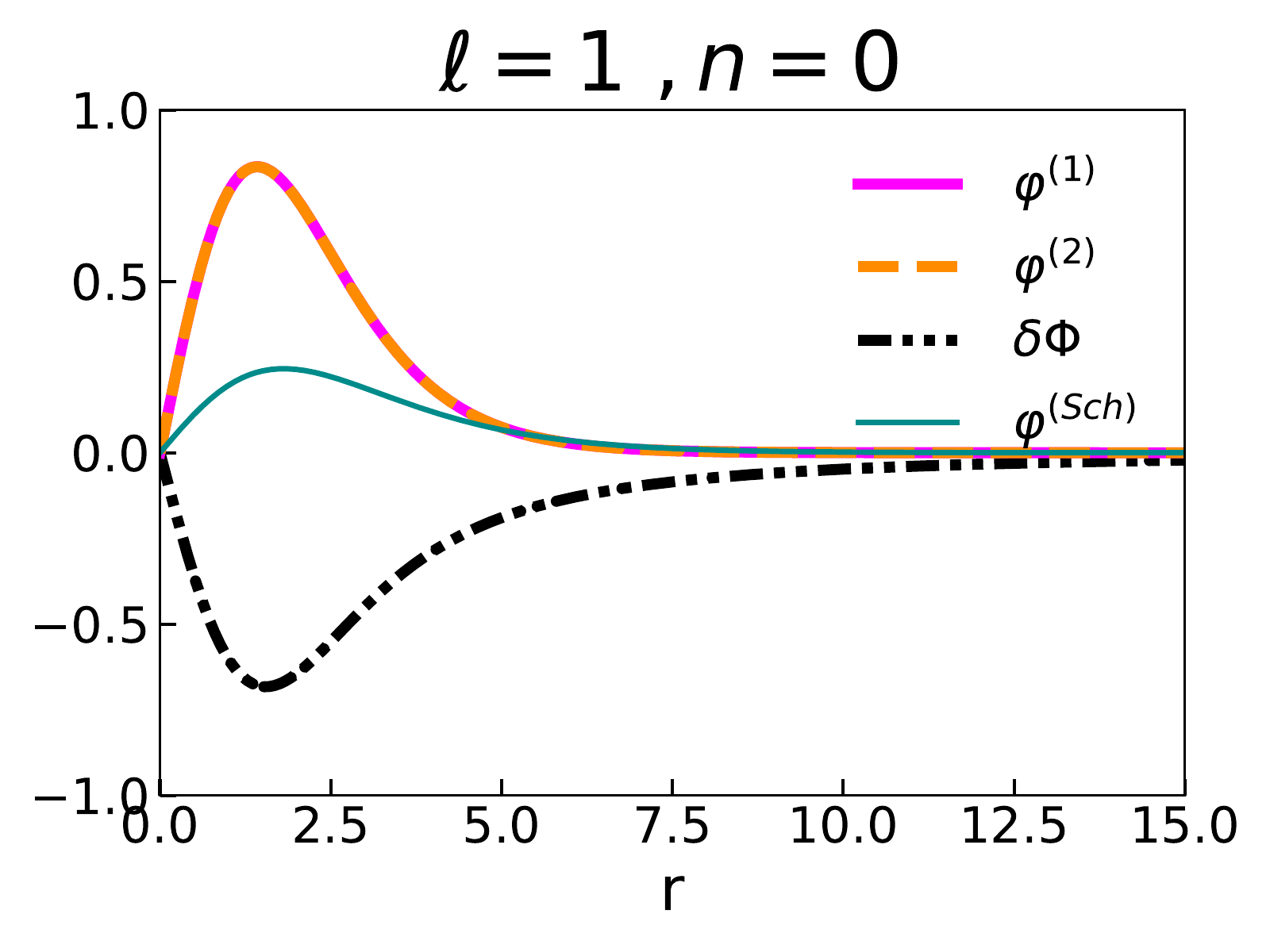}
\includegraphics[width=0.3\columnwidth]{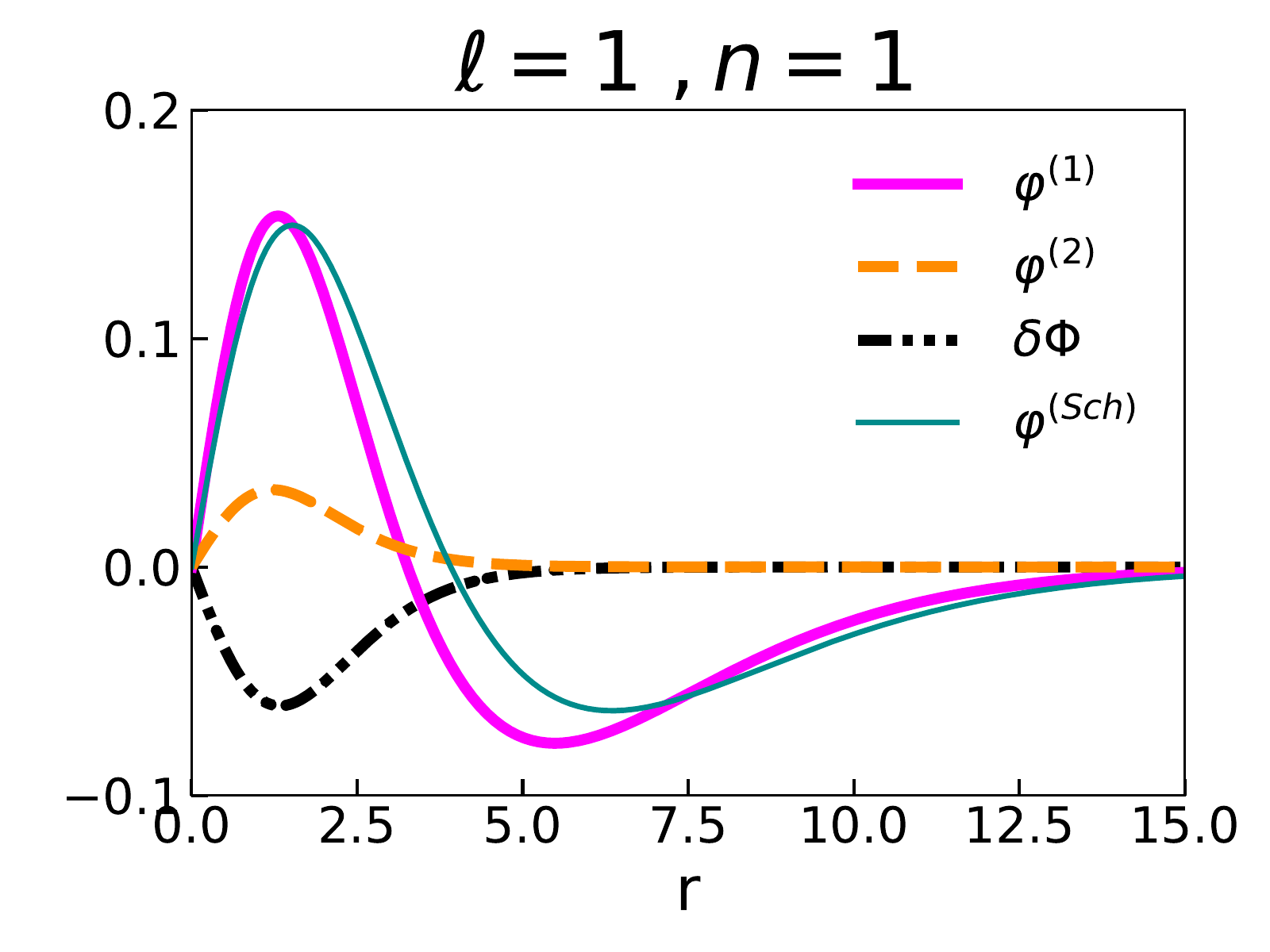}
\includegraphics[width=0.3\columnwidth]{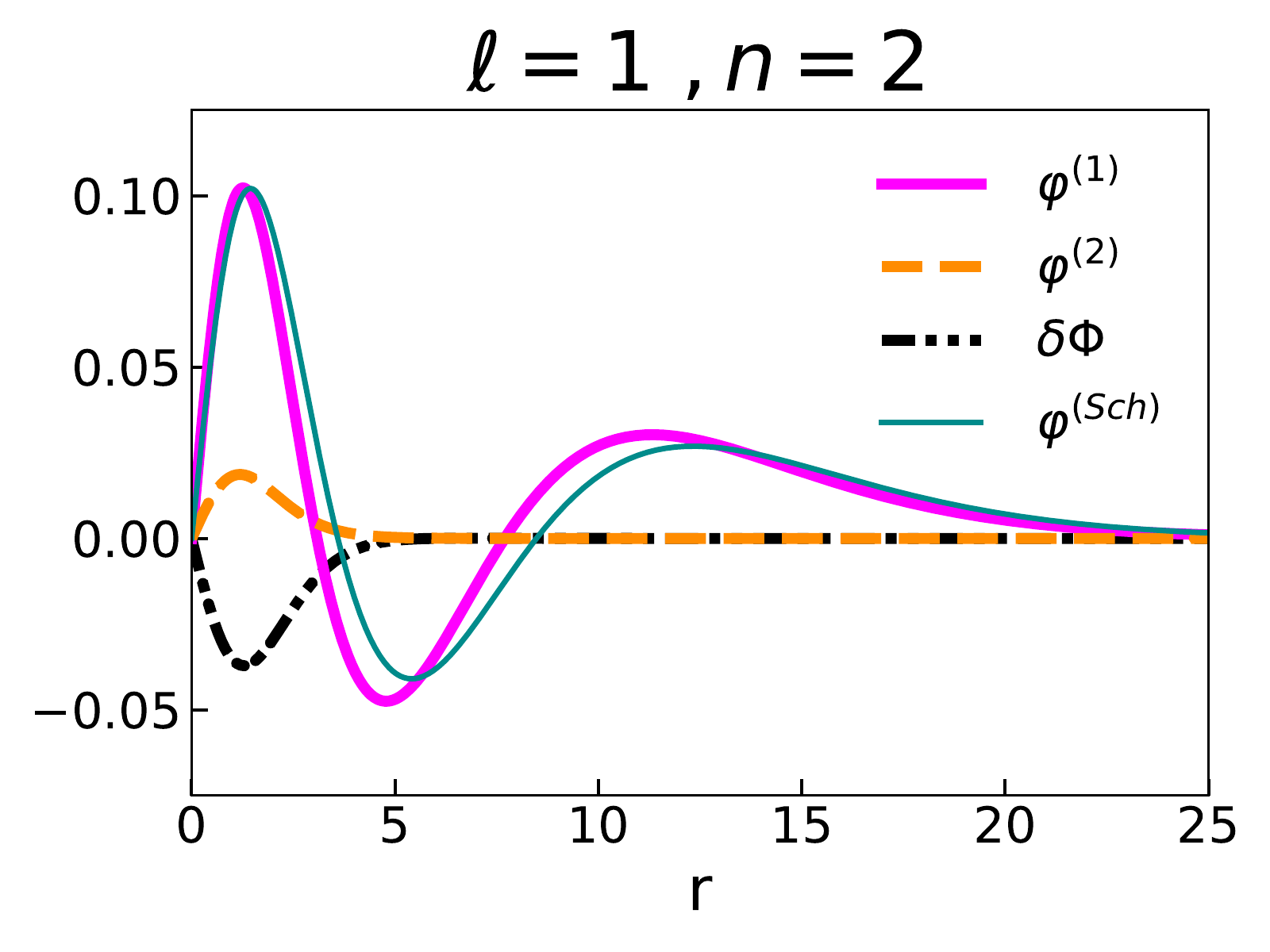}
\includegraphics[width=0.3\columnwidth]{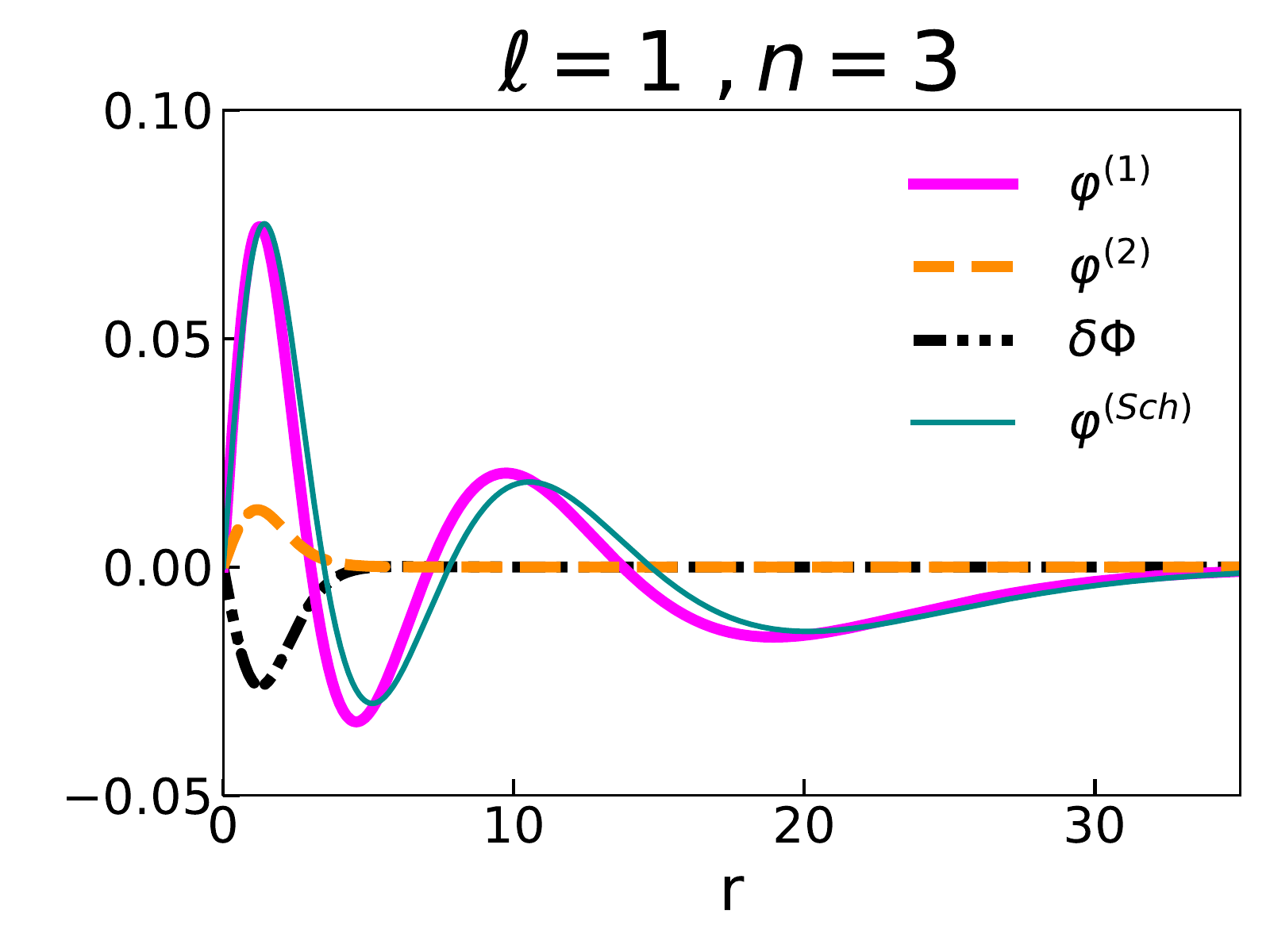}
\includegraphics[width=0.3\columnwidth]{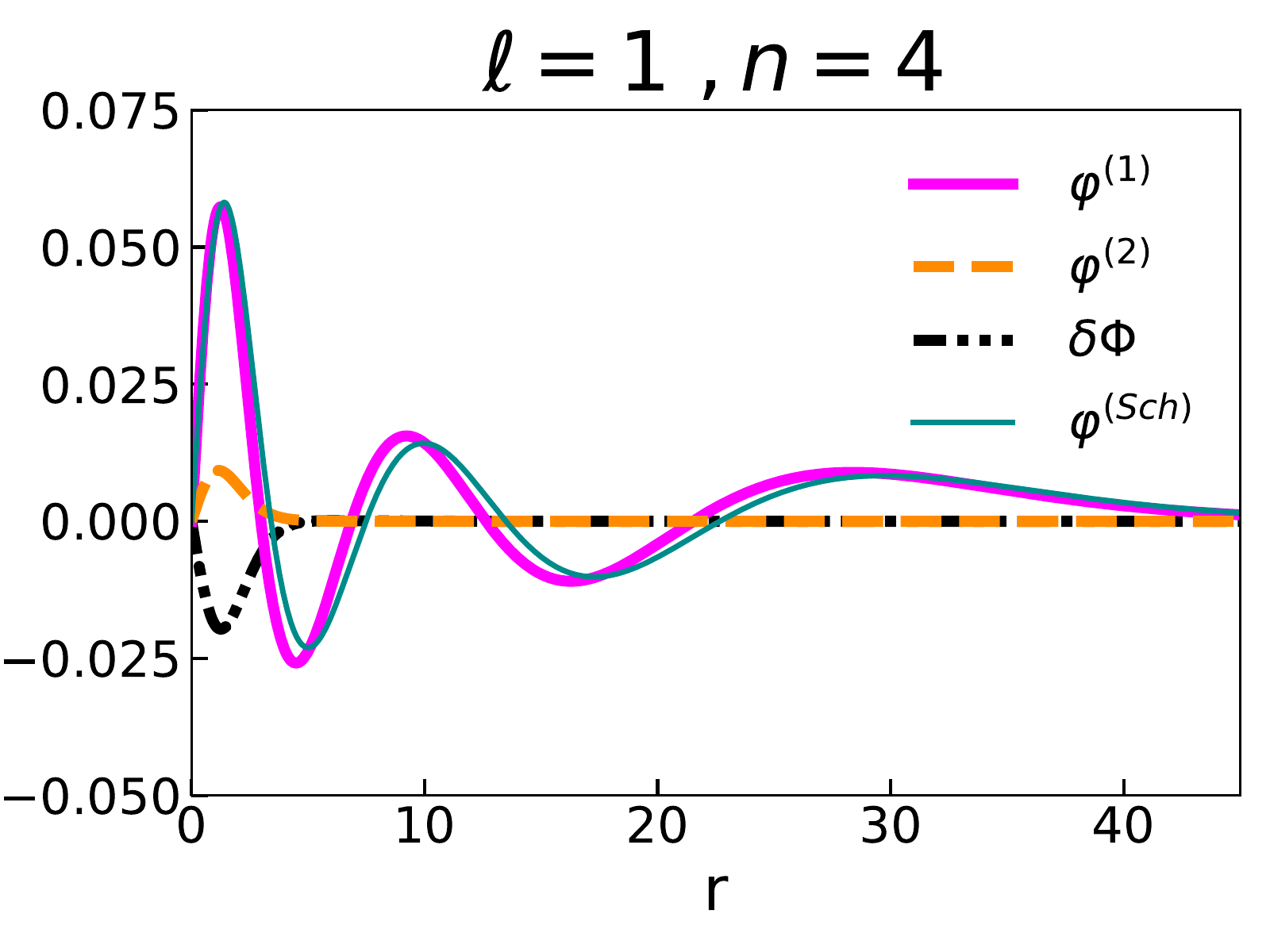}
\\
\includegraphics[width=0.3\columnwidth]{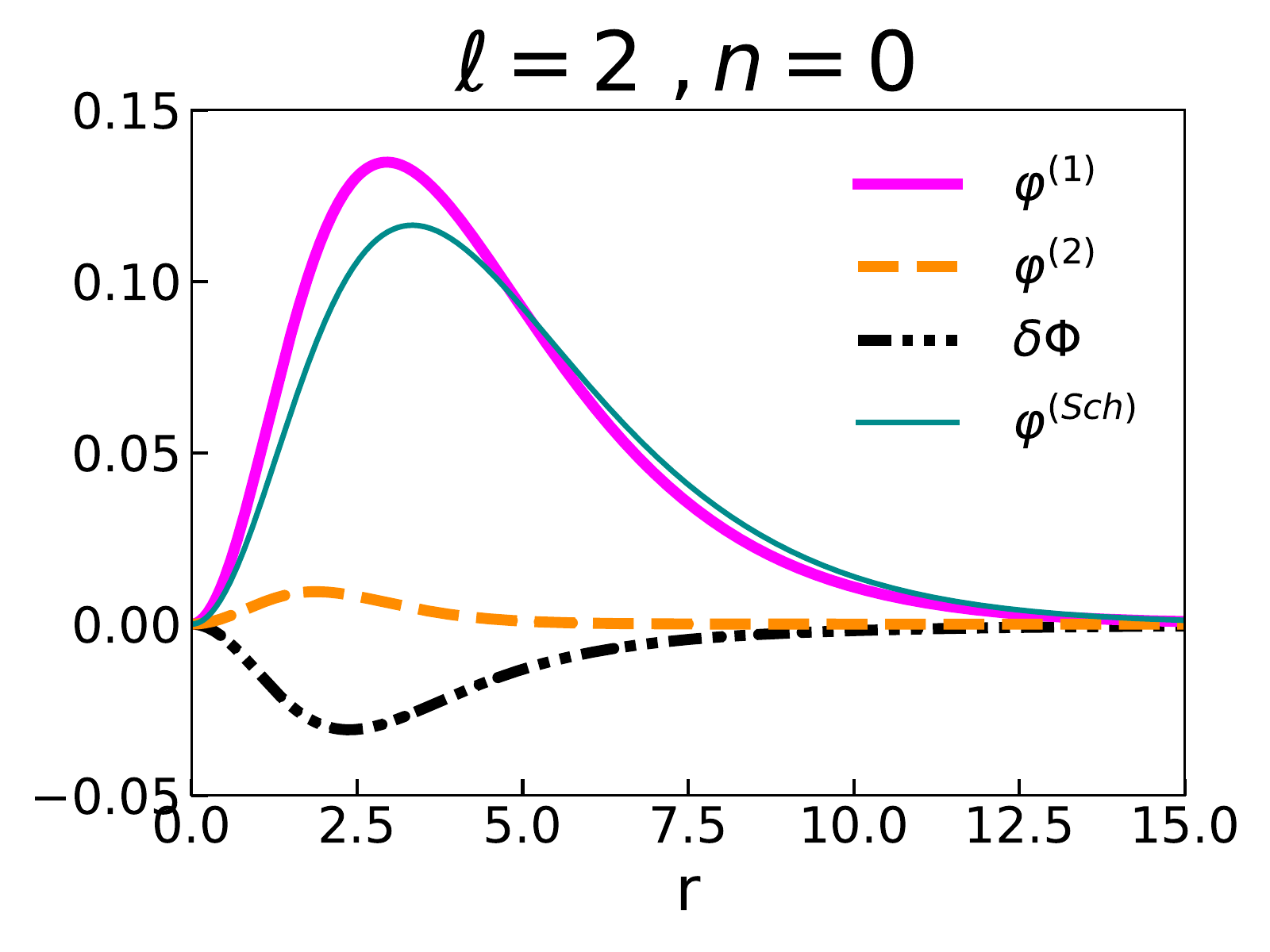}
\includegraphics[width=0.3\columnwidth]{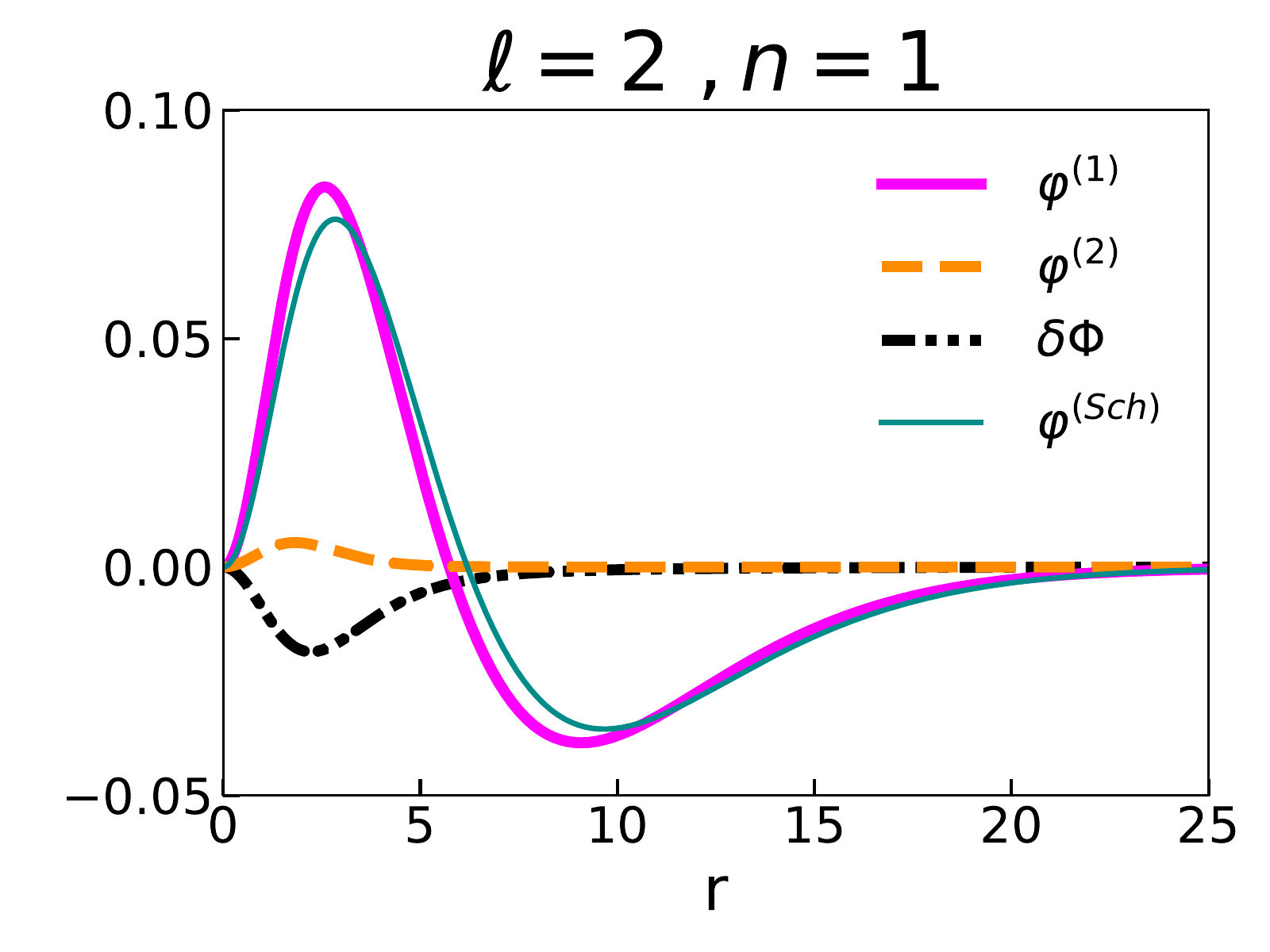}
\includegraphics[width=0.3\columnwidth]{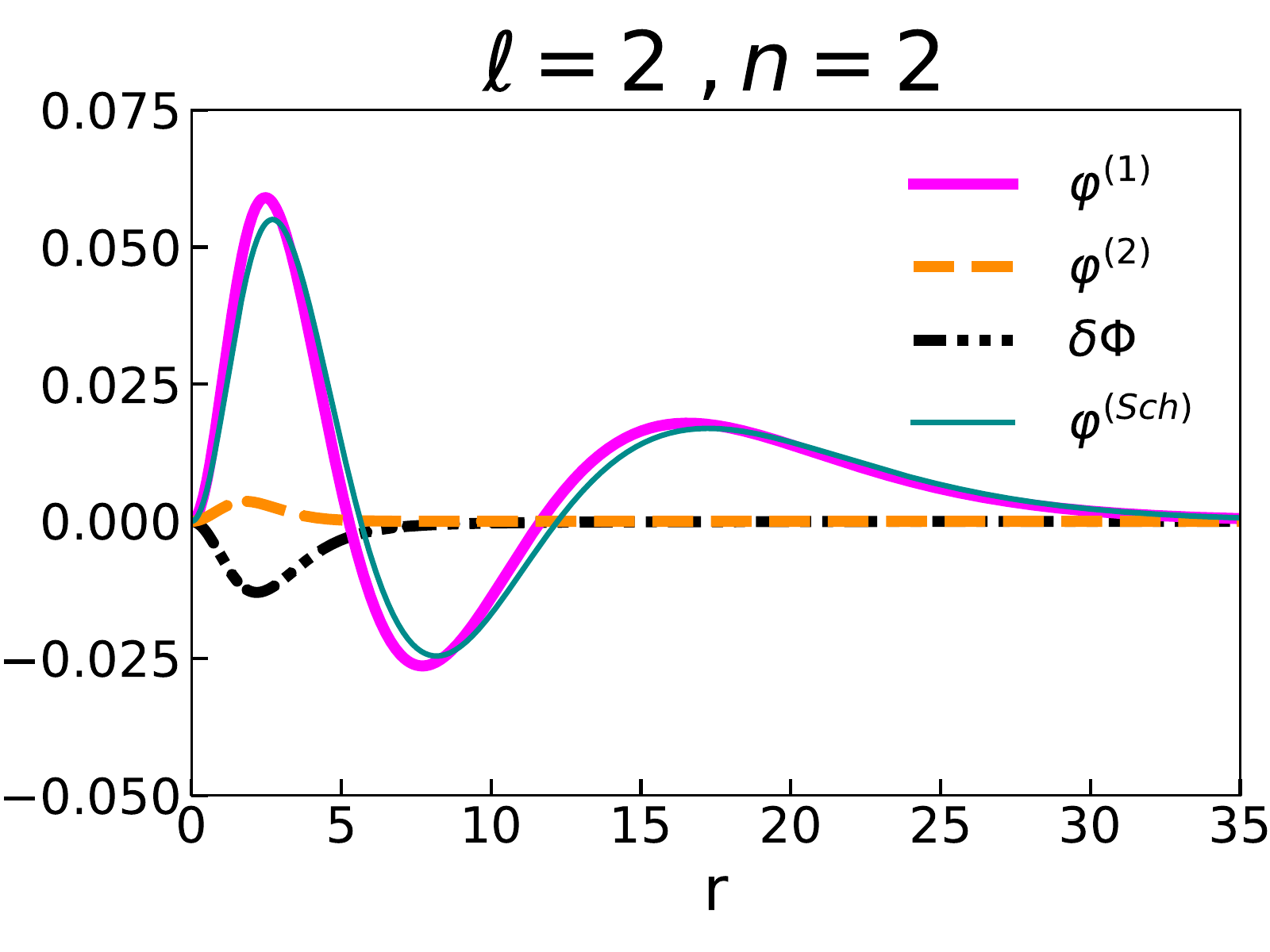}
\\
\includegraphics[width=0.3\columnwidth]{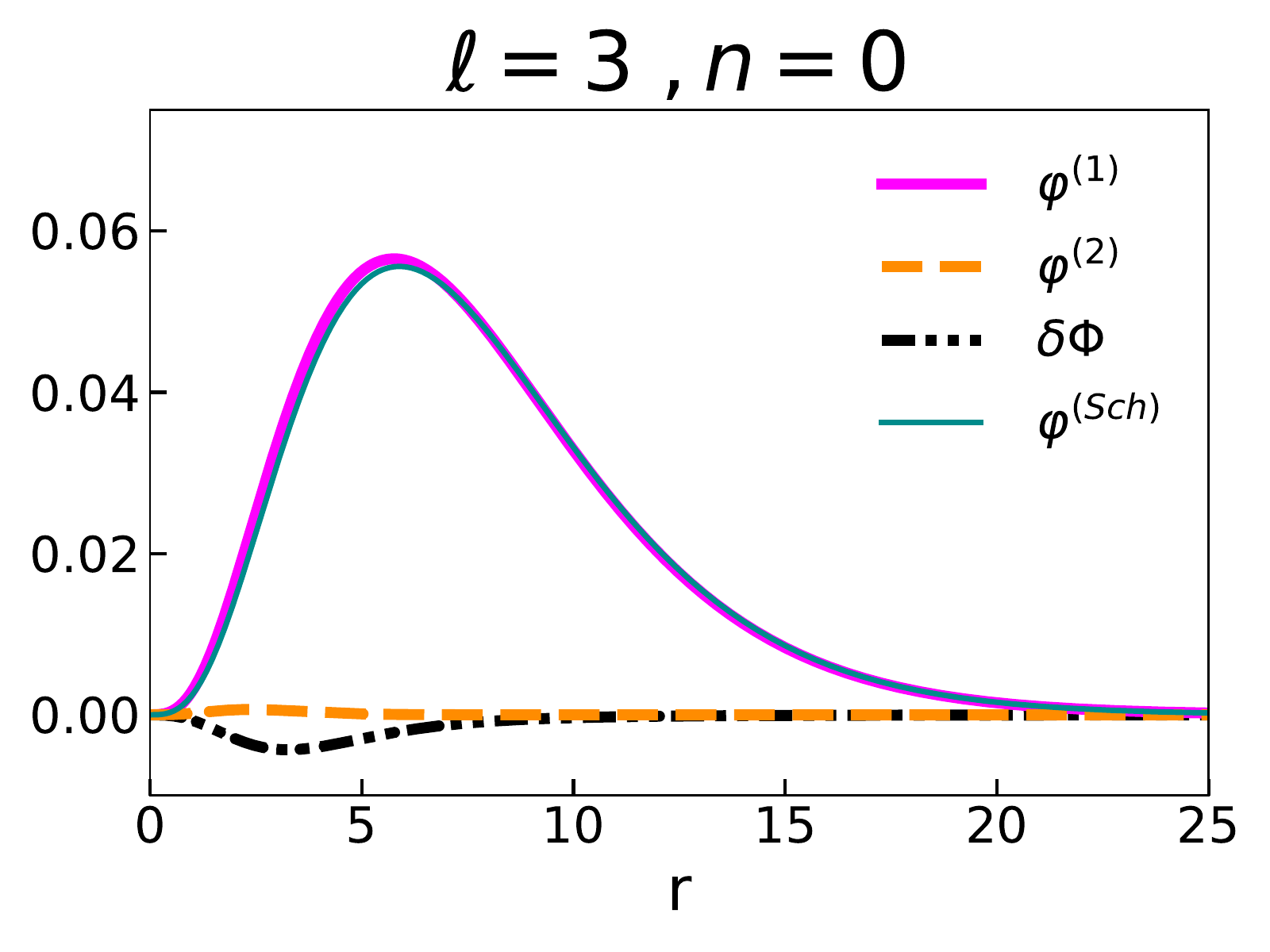}
\includegraphics[width=0.3\columnwidth]{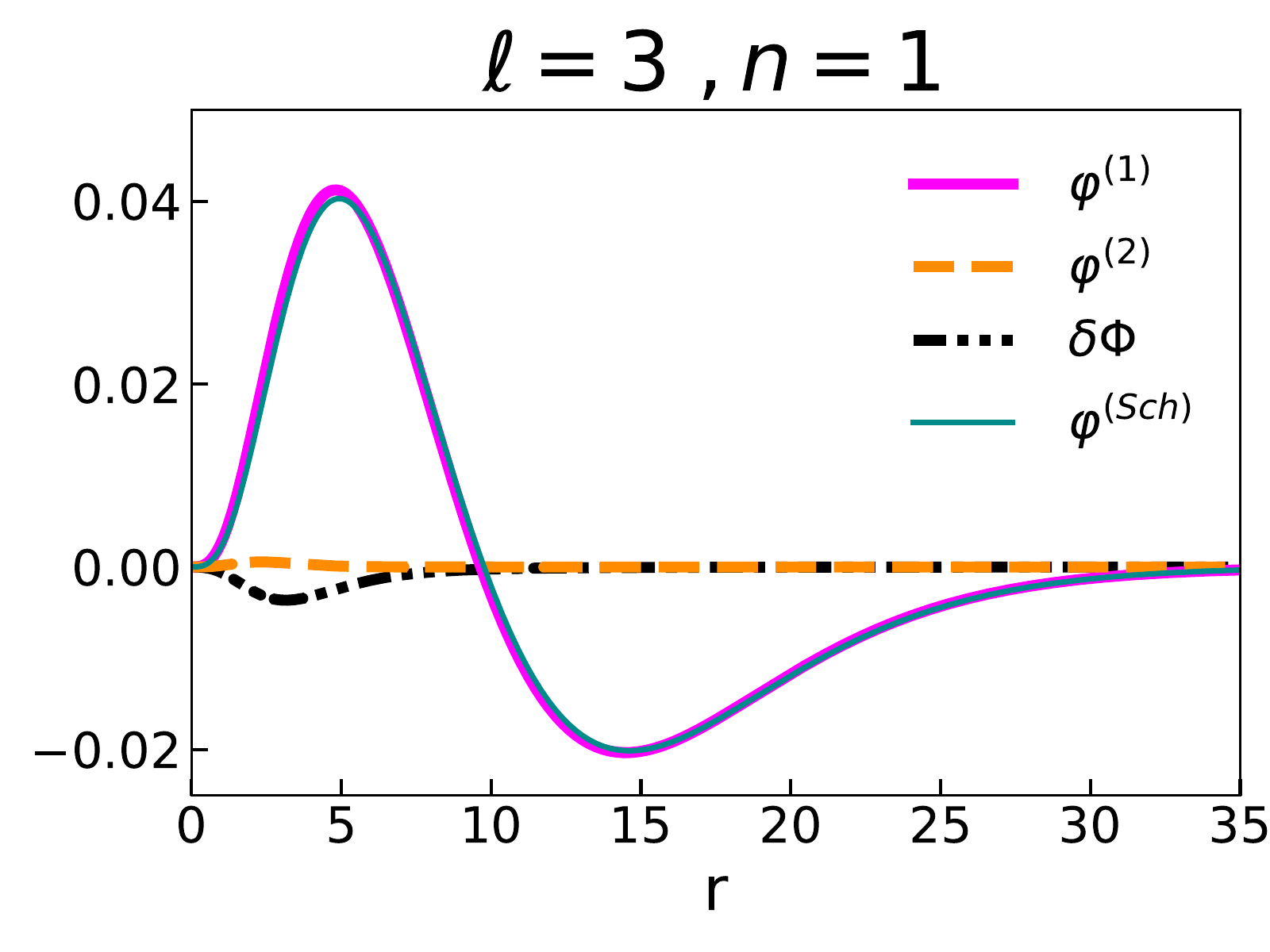}
\includegraphics[width=0.3\columnwidth]{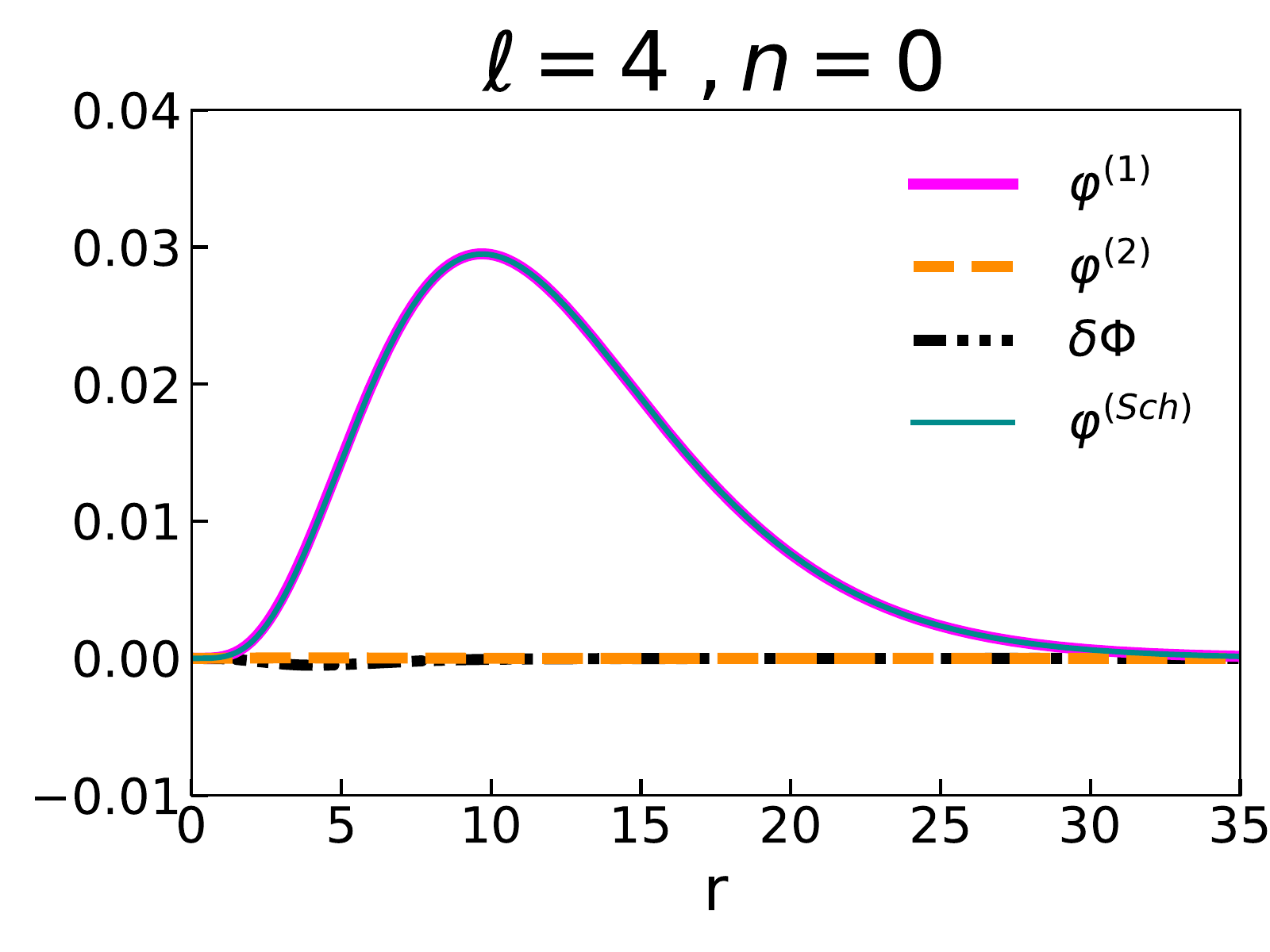}
\caption{The numerical solutions of the linearized
  Schr\"{o}dinger--Poisson (SP)
  system and the Schr\"{o}dinger equation
for $\ell = 0, 1, 2, 3, 4$. Solid magenta lines represent the
positive frequency component of the SP eigenstate, orange dashed
lines --- its negative frequency component, and black dash-dotted
lines the perturbation of the gravitational potential. The turqoise lines
show the Schr\"{o}dinger wavefunctions. 
For each multipole, the first few energy eigenstates are displayed.
For visualization purpose we normalize $\vf_{10}^{(1)}$ to have unit
derivative at $r=0$, which differs by a constant factor from
\cref{eq:groundstate}. 
}
\label{fig:H_wave}
\end{center}
\end{figure}

The numerical results for the wavefunctions and the energy levels are
summarized in \cref{fig:H_wave} 
and \cref{table:eng}. 
We obtain the first few soliton eigenmodes for $\ell = 0, 1, 2, 3, 4$ and 
compare them to the Schr\"{o}dinger solutions. In the last row of the
table we also show the results of perturbative improvement of the
Schr\"odinger approximation, to be discussed in the next subsection. 
Evidently, the positive frequency component of the exact eigenmodes
gets closer to the Schr\"{o}dinger wavefunction for higher levels and
large $\ell$. At the same time, the negative frequency component
becomes negligible. This justifies the replacement of the exact SP
problem by the 
Schr\"{o}dinger approximation with fixed gravitational potential at
high energy levels  
to simplify some computations. 
Note also the degeneracies between the high energy levels with
different $\ell$ and $n$, but the same sum $\ell+n$. They stem from 
the fact that the wavefunctions of highly excited states
are much wider than the soliton (see \cref{fig:H_wave}) and
effectively feel only the Coulomb-like tail of the soliton
gravitational potential ${\Phi_0(r)\simeq -\mu_0/(4\pi r)}$, $r\gg 1$. 

For the low energy levels and small $\ell$, the two solutions of the SP
system and the Schr\"odinger equation are essentially different.   
For instance, the numerical solutions of the energy levels in
\cref{table:eng} confirm that there are monopole and dipole zero
energy modes for  
the exact solutions, but the Schr\"{o}dinger equation has only the 
monopole zero-energy mode.
This is consistent with the fact that the Schr\"{o}dinger equation
explicitly breaks the translational symmetry. We have also checked
that the exact zero modes of the SP system have the expected form
(\ref{eq:groundstate}). 

Our results for the eigenfrequencies in the monopole sector agree
  with the previous studies \cite{Harrison_2003,Guzman:2004wj}  
 within 
  $O(10^{-3})$.\footnote{Ref.~\cite{Harrison_2003} uses
  a different normalization of the soliton, so their values must be
  rescaled by a factor $8.49$ for comparison.}

\begin{table}[ht]
\centering
\begin{tabular}{ |c | *{5}{|c} | } 
\hline
      &   \multicolumn{5}{c|}{$\ell = 0 $}  \\
\cline{2-6}
 & $n = 0 $  &  $n = 1$ & $n = 2$  & $n = 3$  & $n = 4$  \\
\hline
\hline
SP eq.
& 0  & 0.290   & 0.513 & 0.585 & 0.621  \\
\hline
Sch. eq. 
& 0  &  0.394 &  0.533 & 0.593 & 0.625  \\
\hline
\hline
Pert. th.
& --  &  0.334 &  0.514 & 0.586 & 0.621  \\
\hline
\end{tabular}
\\
\vspace*{0.5 cm}
\begin{tabular}{ |c | *{5}{|c} | } 
\hline
      &   \multicolumn{5}{c|}{$\ell = 1 $}  \\
\cline{2-6}
 & $n = 0 $  &  $n = 1$ & $n = 2$  & $n = 3$  & $n = 4$  \\
\hline
\hline
SP eq.
& 0  &  0.472    &  0.568 & 0.612 & 0.636  \\
\hline
Sch. eq. 
& 0.287  &  0.495 & 0.576  & 0.616  & 0.638 \\
\hline
\hline
Pert. th. 
& 0.177 &  0.473    &  0.568 & 0.612 & 0.636  \\
\hline
\end{tabular}
\\
\vspace*{0.5 cm}
\begin{tabular}{ |c | *{3}{|c} ||  c| c|| c|} 
\hline
      &   \multicolumn{3}{c||}{$\ell = 2 $} & \multicolumn{2}{c||}{$\ell = 3 $}  & $\ell = 4$  \\
\cline{2-7}
 & $n = 0 $  &  $n = 1$ & $n = 2$  & $n = 0$  & $n = 1$ &  $n = 0 $  \\
\hline
\hline
SP eq.
&  0.450  &  0.558  &  0.607  & 0.559  & 0.607  &   0.607 \\
\hline
Sch. eq. 
& 0.466 &  0.564 &  0.610  & 0.560 & 0.607 &   0.607 \\
\hline
\hline
Pert. th.
&  0.453  &  0.559  &  0.607  & 0.559  & 0.607  &   0.607 \\
\hline
\end{tabular}
\caption{The energy levels in the background of the
  standard soliton obtained using the full 
 Schr\"{o}dinger--Poisson (SP) system and their estimates from the 
Schr\"{o}dinger (Sch.) approximation.
The last row shows the Schr\"{o}dinger eigenvalues improved using
perturbation theory described in \cref{subsec:pert}. 
}
\label{table:eng}
\end{table}

\subsection{Perturbation theory}
\label{subsec:pert}

We have seen above that the eigenvalues of the full SP system and the
Schr\"odinger approximation approach each other for high energy
levels. Since the single Schr\"odinger equation is significantly
easier to solve than the SP system, it motivates us to look for the
solution of the latter perturbatively, starting from the Schr\"odinger
wavefunction at the zeroth order. This is similar to the perturbation
theory in quantum mechanics with $H_0$ playing the role of the
unperturbed Hamiltonian and the terms $V$ in
\cref{eq:pertEqMatrix2}
treated
as the perturbation. 

Since the eigenfunctions of $H_0$ form a complete basis, we can expand
the positive and negative frequency components of the exact eigenstate
as 
\begin{equation} 
   \varphi_a^{(1)}({\bf x}) 
= \sum_b  \beta^{(1)}_{ab} \, \varphi_b^{\rm(Sch)}  ({\bf x})   \, ,
      \quad 
   \varphi_a^{(2)}({\bf x})
= \sum_b  \beta^{(2)}_{ab} \, \varphi_b^{\rm(Sch)}  ({\bf x})   \, .
   \label{eq:C12}
\end{equation} 
We insert this expansion into 
\cref{eq:pertEqMatrix2}, multiply by  
${\varphi_c^{\rm (Sch)}}^*({\bf x})$, integrate over ${\bf x}$ and use
the orthonormality of the Schr\"odinger basis. This gives 
the conditions for the coefficients, 
\bseq
\label{eq:C}
\begin{eqnarray}
   && \beta_{ac}^{(1)} \, (\epsilon^{\rm (Sch)}_c - \epsilon_a ) +   
         \sum_b \left( \beta_{ab}^{(1)} + \beta_{ab}^{(2)} \right) \, 
V_{cb}^{\rm (Sch)} = 0   \, , 
   \label{eq:C1}
   \\
   && \beta_{ac}^{(2)} \, (\epsilon^{\rm (Sch)}_c + \epsilon_a ) +   
         \sum_b \left( \beta_{ab}^{(1)} + \beta_{ab}^{(2)} \right) \, 
V_{cb}^{\rm (Sch)} = 0   \, ,  
   \label{eq:C2}
\end{eqnarray}
\eseq
with the matrix elements
\be
 V_{cb}^{\rm (Sch)} = \int d^3 {\bf x}  \, {\varphi_c^{\rm (Sch)}}^* ( {\bf
   x}) \, V \, \varphi_b^{\rm (Sch)} ( {\bf x})   
         =   \int d^3 {\bf x}  \, {\varphi_c^{\rm (Sch)}}^*  ( {\bf x}) \,
         \chi (r) \,   \frac{1} {\Delta} \left(\chi(r)  
      \varphi_b^{\rm (Sch)} ( {\bf x})  \right)
      \, . 
\ee
Decomposing  $\varphi_a^{\rm (Sch)} ({\bf x})$ 
into the spherical harmonics and the radial part
(cf. \cref{eq:decompvarphi}) we can write them as 
\be
\begin{split}
   V_{\ell, n, m\,;\,\ell', n', m'}
   = \delta_{\ell\ell'}
      \delta_{mm'}\times
\left(-\frac{ 2} {\pi}\right) \int_0^\infty dk& \, 
      \left( \int_0^\infty dr \, r^2 \, j_{\ell} ( k r ) \, 
{\varphi^{\rm(Sch)}_{\ell n}}^* ( r) \, \chi ( r) \right)   
      \\
      & \times \left( \int_0^\infty dr' \, r'^2 \, j_{\ell} ( k r' ) 
\, \varphi_{\ell n'}^{\rm(Sch)} ( r' ) \, \chi ( r') \right) \, ,
\end{split}
\ee
where $j_\ell(z)$ are the spherical Bessel functions and we have used
the identity 
\be
 \int d \hat {\bf x} \, e^{i {\bf k} \cdot {\bf x} } \, Y_{\ell m }
(\hat{\bf x} ) 
      = i^\ell 4 \pi \, j_{\ell} ( kr ) Y_{\ell m}^* ( \hat{\bf k})\;.
\ee
By subtracting eq.~(\ref{eq:C2}) from eq.~(\ref{eq:C1}), we find a
simple relation,\footnote{This relation does not apply to the monopole
  ground state, for which $\epsilon_{00}=\epsilon_{00}^{\rm (Sch)}=0$.
In this case, however, we do not need perturbation theory since the
corresponding mode is known exactly,
$\vf_{00}^{(1)}=\beta_{00}\vf_{00}^{\rm (Sch)}=\chi$,
$\vf_{00}^{(2)}=-\beta_{00}\vf_{00}^{\rm (Sch)}=-\chi$, where the
coefficient $\beta_{00}$ is due to different normalization of the
Schr\"odinger and exact modes.  
}  
\begin{equation}
   \beta_{ac}^{(2)} =  \frac{\epsilon_c^{\rm (Sch)} - \epsilon_a  } 
{\epsilon_c^{\rm (Sch)} + \epsilon_a} \beta_{ac}^{(1)} \ .
\end{equation}
Inserting the above relation back to eqs.~(\ref{eq:C}) we find,
\bseq
\label{eq:Cmat}
\begin{eqnarray} 
   && \beta_{ac}^{(1)} \, \left(\epsilon_c^{\rm (Sch)}  - \epsilon_a\right) +   
         \sum_b  \frac{ 2 \epsilon_b^{\rm (Sch)}}
{\epsilon_b^{\rm (Sch)}+\epsilon_a}   \, \beta_{ab}^{(1)}  \,
V_{cb}^{\rm (Sch)} = 0   \, , 
   \label{eq:C1mat}
   \\
   && \beta_{ac}^{(2)} \, \left(\epsilon_c^{\rm (Sch)}  + \epsilon_a\right) +   
         \sum_b  \frac{ 2 \epsilon_b^{\rm (Sch)}}
{\epsilon_b^{\rm (Sch)}-\epsilon_a}   \, \beta_{ab}^{(2)}  \,
V_{cb}^{\rm (Sch)} = 0   \, .
   \label{eq:C2mat}
\end{eqnarray} 
\eseq

Until this point, all the formulas have been exact. 
Let us now assume that the matrix elements $V_{cb}^{\rm (Sch)}$ are
small. In the leading order we set them to zero in
eqs.~(\ref{eq:Cmat}) and obtain that the only solution is\footnote{We
  look for solutions with $\epsilon_a\geq 0$.}
\be
\label{eq:PT0}
\epsilon_a=\epsilon_a^{\rm (Sch)}\,,~~~~~~
\beta_{ac}^{(1)}=\delta_{ac}\,,~~~~~\beta_{ac}^{(2)}=0\;.
\ee
As expected, at this order the negative frequency component vanishes,
while the positive frequency one coincides with the Schr\"odinger
wavefunction. 

The next step is to take into account the corrections linear in
$V$. Using (\ref{eq:PT0}) in the terms in eqs.~(\ref{eq:Cmat}) that
are already proportional to $V$, we find at the next-to-leading order, 
\bseq
\label{eq:PT1}
\begin{align}
\label{eq:EPT1}
&\epsilon_a=\epsilon_a^{\rm (Sch)}+V_{aa}^{\rm(Sch)}\,,\\
\label{eq:beta1PT1}
&\beta_{aa}^{(1)}=1\,,~~~~~
\beta_{ac}^{(1)}=\frac{V_{ca}^{\rm (Sch)}}{\epsilon_a^{\rm (Sch)}
-\epsilon_c^{\rm (Sch)}}~~\text{for} ~~c\neq a\,,\\
\label{eq:beta2PT1}
&\beta_{ac}^{(2)}=-\frac{V_{ca}^{\rm (Sch)}}{\epsilon_a^{\rm (Sch)}
+\epsilon_c^{\rm (Sch)}}
\;.
\end{align}
\eseq
The results of the numerical evaluation of the energy levels using
\cref{eq:EPT1} are presented in the last row of \cref{table:eng}. 
They are in excellent agreement with the exact eigenvalues 
for almost all the modes, 
except the lowest levels $(\ell = 0 , n = 1)$ and $(\ell = 1 , n = 0)
$. Using eqs.~(\ref{eq:beta1PT1}), (\ref{eq:beta2PT1}) to find the
wavefunctions is more
challenging since the expansions (\ref{eq:C12}) involve summation over
infinitely many Schr\"odinger modes. We have found that the sums
converge rather slowly and we do not attempt to evaluate them here.

\section{Soliton Spectroscopy with 3D Simulations}
\label{sec:wavesimulation}

In this section we observe how the energy levels of the soliton bound
states manifest themselves in the spectrum of soliton perturbations
in dynamical wave simulations on a 3-dimensional grid.
We inject $\ell = 0, 1, 2$ perturbations separately in the simulations
and perform the frequency analysis of the density field at several
points in space. Its power spectrum contains resonant lines
corresponding to the soliton eigenmodes with 
the three orbital angular momenta~$\ell$. 
The resonant frequencies 
excellently match the prediction given by the theoretical and
numerical studies before. 

We follow the same approach and conventions as
ref.~\cite{Chan:2022bkz} to perform the wave simulations. A soliton
wavefunction is  
introduced in a cubic lattice of size ${\cal N} = 256$ with periodic
boundary conditions. The axion field is evolved by solving 
the dimensionless SP equations (we set $4\pi G=m=1$),
\bseq
\label{eq:dimensionlessEq}
\begin{align}
\label{eqSchrSim}
&i\d_t \psi +\frac{1}{2} \Delta\psi - \Phi \,
\psi=0\;, \\ 
\label{eqPoisSim}
&\Delta\Phi=|\psi|^2\;.
\end{align}
\eseq
We further fix the lattice spacing to unity, ${\rm d} x = 1$, by the remaining
scaling symmetry of  
the Schr\"{o}dinger-Poisson equations.
  
We use the soliton background shown in \cref{fig:standsol} as an input
for the lattice initial condition, rescaling it by a factor ${\cal S}
= 0.428$ in order to achieve a better spatial and temporal
resolution. The parameters of the soliton are fixed accordingly: the
peak density $\rho_c = 0.0337$, the radius at half peak density $r_{1/2} = 3.03$, 
the total mass $M_s = 11.1$, and the binding energy $\E_s = -0.127$. 
The rescaling also applies to all the energy levels of the soliton
perturbations, some of which are listed in \cref{table:eng}: 
An energy level is rescaled to ${\cal S}^2 \epsilon_{ \ell n}$ in the
simulation setup.  

To analyze the soliton spectrum, we follow ref.~\cite{Guzman:2018bmo}
and introduce a perturbation $\delta \psi$ with a fixed
orbital angular momentum $\ell$  
of the form
\begin{equation}
   \delta \psi ( r, \theta, \phi ) = A_{\ell 0 } \, \exp \big( - ( r -
   r_p )^2 / \sigma_p^2 \big ) \, Y_{\ell 0} ( \theta)   \ .
\end{equation}
The perturbation represents a wavepacket of width $\sigma_p$ in the
radial direction and centered at $r_p$. Its angular dependence is a
pure spherical harmonic $Y_{\ell m}$. 
We choose $m = 0$, so that the perturbation is axially symmetric (no
$\phi$-dependence).   
In the simulations, 
we take $r_p = 20 $, $\sigma_p = 1$ and $A_{\ell 0} =
0.01,\,0.005,\,0.0025$ for $\ell=0,\,1,\,2$, which
perturbs the soliton mass by no more than
$\sim 0.5 \%$. 

We evolve a soliton and its perturbation on the lattice using the
leapfrog integration algorithm \cite{birdsall2018plasma,ChanEtal18},  
\begin{equation}
\label{DKD}
   \psi(t+ {\rm d} t, {\bf x} )= \e^{i \, \Delta \, {\rm d} t/4} 
\cdot \e^{-i \, \Phi(t+{\rm d} t/2, {\bf x}  )  \,  
      {\rm d} t } 
            \cdot \e^{i \, \Delta \, {\rm d} t/4}  \, \psi(t, {\bf x}  )  \, .
\end{equation}
The time evolution with the operator $\e^{i \, \Delta \, {\rm d} t/4}$ is computed 
in the momentum space, while the phase $\e^{-i \, \Phi(t+{\rm d} t/2, {\bf x}  )}$
is evaluated in real space. 
We choose ${\rm d} t =  1/ (2 \pi )$ and the simulation time $t_{\rm max}
= 80,000 \, {\rm d}t $. 
In order to perform the spectroscopic  analysis of the system, we
Fourier transform the density $\rho (t, {\bf x}) = |\psi( t,{\bf
  x})|^2$  
into its frequency 
space $\rho( \omega ,{\bf x})$.
The resolution in frequency space is determined by the total
simulation time, ${\rm d}\omega=\pi/t_{\rm max}\sim 4\cdot 10^{-5}$.  

The power spectra for $\ell = 0,1 ,2 $ perturbations $P (\omega, {\bf
  x}) = | \rho( \omega ,{\bf x}) |^2 $ are presented in  
figs.~\ref{fig:l0},~\ref{fig:l1},~\ref{fig:l2}, in which 
the four panels show the power spectra measured at ${\bf x} = ( 0, 0, 0 ) $, $(
5, 0, 0) $, $(0,  5, 0)$ and $(0, 0, 5)$ in   
Cartesian coordinates. The visible resonant 
peaks are consistent with energy eigenvalues predicted in the previous study.
The summary and analysis of the three perturbations are given in order:

 \begin{figure}[h]
\begin{center}
 \includegraphics[width=1.0\textwidth]{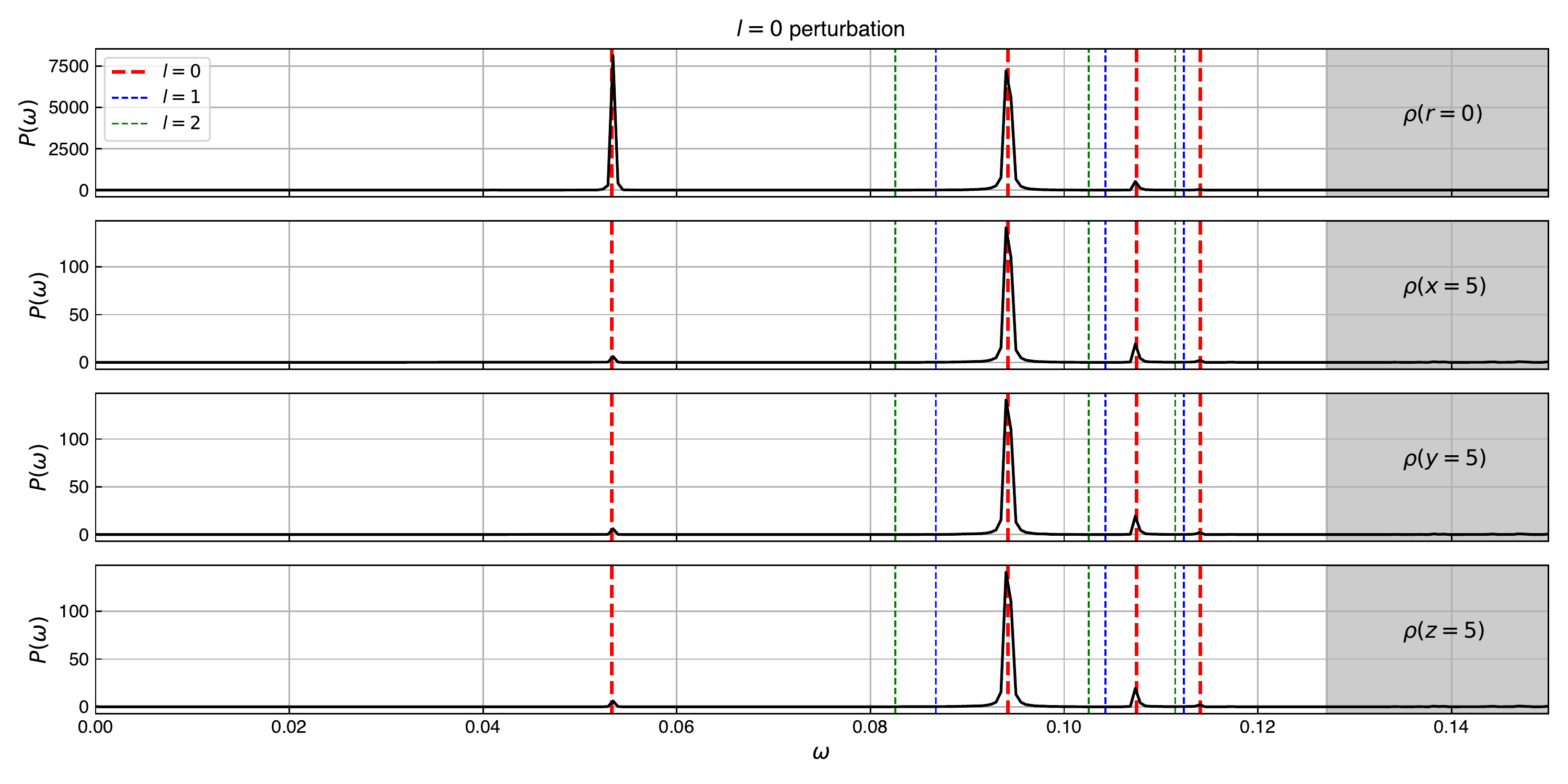}
\caption{The power spectrum of $\ell = 0 $ perturbation (black lines) at four locations: $r = 0 $, $x = 5$, $y=5$, and $z=5$.
The vertical dotted red lines correspond to the predicted binding energy of $\ell = 0 $ modes and $n = 1$ to $4$;
the vertical dotted blue lines correspond to the binding energy of $\ell = 1 $ modes and $n = 1$ to $3$;
the vertically dotted green lines correspond to the predicted energy levels of $\ell =2$ modes and $n = 0$ to $2$.
The grey-shaded region represents the start of the continuum spectrum.
The same color coding applies to the power spectrum analysis of $\ell = 1$ and $\ell =2$ shown in \cref{fig:l1,fig:l2}.
}
\label{fig:l0}
\end{center}
\end{figure}

\subsubsection*{$\ell = 0$ perturbation:}
In \cref{fig:l0}, the vertical dotted red lines correspond to the predicted energy levels of $\ell = 0 $ modes with $n = 1$ to $4$
(see \cref{table:eng}), rescaled to the simulation units. 
In each panel ($r = 0 $, $x = 5$, $y=5$ and $z=5$), there are four
visible peaks, though the fourth one is hardly seen.  
The locations of the peaks match well with the vertical red lines.
The fundamental $n=1$ mode is prominent in the soliton center ($r=0$),
but gets weak at the distance $r=5$ because its wavefunction quickly
decreases outside the soliton. Also, it is less excited than the $n=2$
mode since the injected perturbation is localized far away from the
center. As a consequence, the most prominent mode at $x=5$,
$y=5$ and $z=5$ is the second mode 
$n=2$. Note that the amplitudes of the $n=2$ peaks are
similar in all directions which is consistent with the monopole nature
of the perturbation.

\subsubsection*{$\ell = 1$ perturbation:}
\begin{figure}[h]
\begin{center}
 \includegraphics[width=1.0\textwidth]{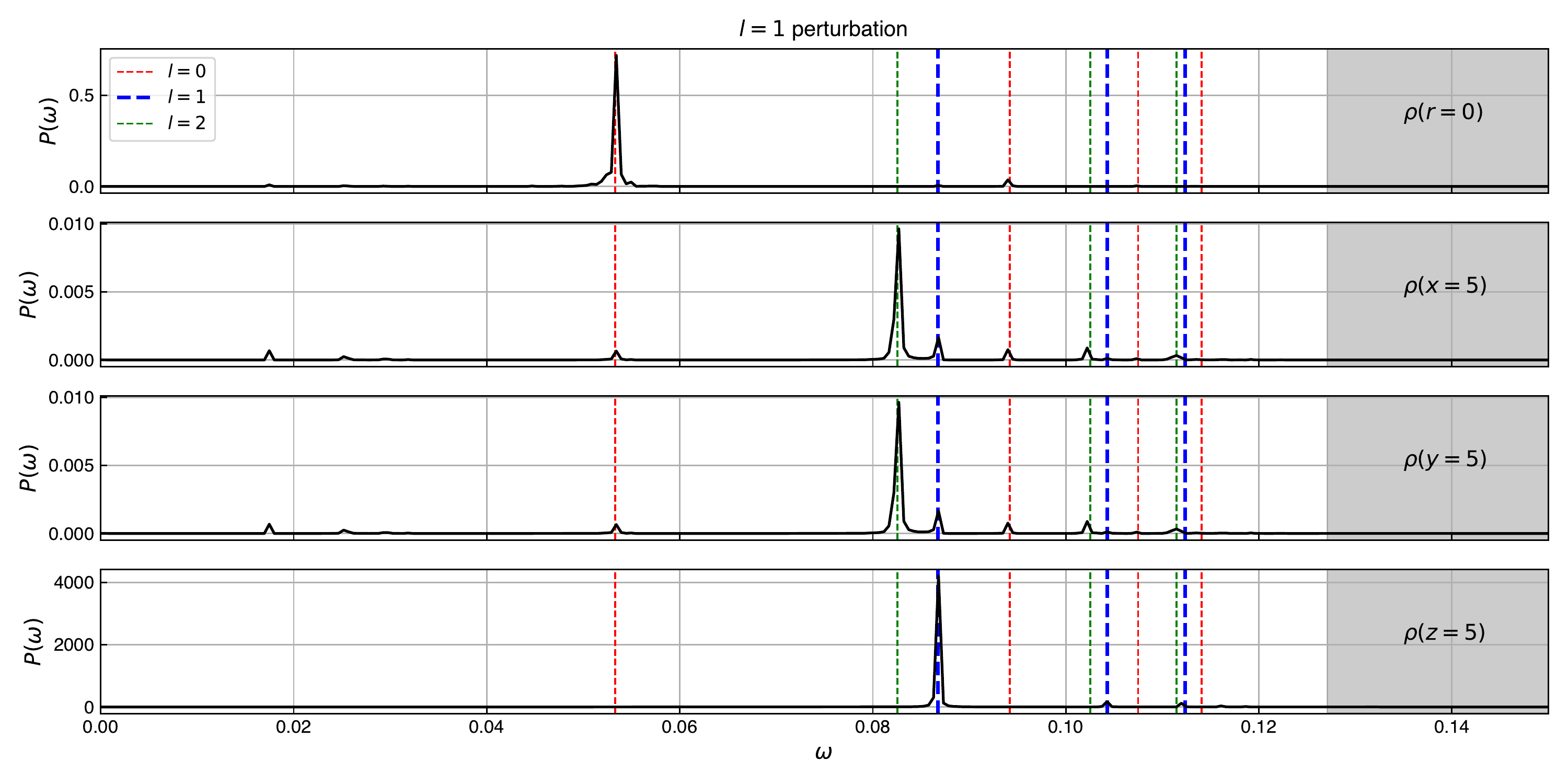}
\caption{The power spectrum of $\ell = 1 $ perturbation (black lines) at four locations: $r = 0 $, $x = 5$, $y=5$ and $z=5$.
The same color coding as \cref{fig:l0} is used. Notice the vast
difference in vertical scale in different panels.
}
\label{fig:l1}
\end{center}
\end{figure}

In \cref{fig:l1}, the vertical dotted blue lines correspond to the
predicted energy levels of $\ell = 1$ modes and $n = 1$ to $3$.
We see a prominent peak corresponding to the $(\ell, n)=(1,1)$ 
mode in the
$z=5$ panel, together with much smaller peaks corresponding to
$(\ell,n)=(1,2)$ and $(1,3)$ modes. The peaks are essentially absent
in all other panels --- notice the much smaller vertical
scale in these panels. This is consistent with the structure of the
injected perturbation which is proportional to  
$Y_{1,0} ( \theta)$ and vanishes in the $z=0$ plane. 

At $r=0$ and $x=5$, $y=5$ we still observe some weak resonant lines
corresponding to $\ell=0$ and $\ell=2$ modes. This may be due to the
mode mixing caused by the discretization and boundary effects from the
lattice.  

Ref.~\cite{Guzman:2018bmo} reports a resonant peak in the dipole
sector at very low frequency, about $20$ times lower than the
frequency of the $(\ell,n)=(1,1)$ mode. We do not see such peak in our
simulations. We believe that its appearance is due to an explicit
breaking of the translation invariance by the code used in
\cite{Guzman:2018bmo} which discretizes the equations in cylindrical
coordinates and thus introduces a preferred origin at the level of the
lattice. As a results, the zero dipole mode associated to this
symmetry acquires a small non-zero frequency. Our code, on the other
hand, preserves exact lattice translation invariance thanks to the use
of Cartesian coordinates and periodic boundary conditions, so the
zero frequency of the translational mode remains protected. It is worth
stating, however, that in a physical situation of a soliton embedded
in a dark matter halo, the translation invariance is expected
to be broken by the gravitational potential of the halo, so that the
translational mode will have non-zero frequency. The precise value of
the frequency will depend on the halo density and its profile.

\subsubsection*{$\ell = 2$ perturbation:}
\begin{figure}[h]
\begin{center}
 \includegraphics[width=1.0\textwidth]{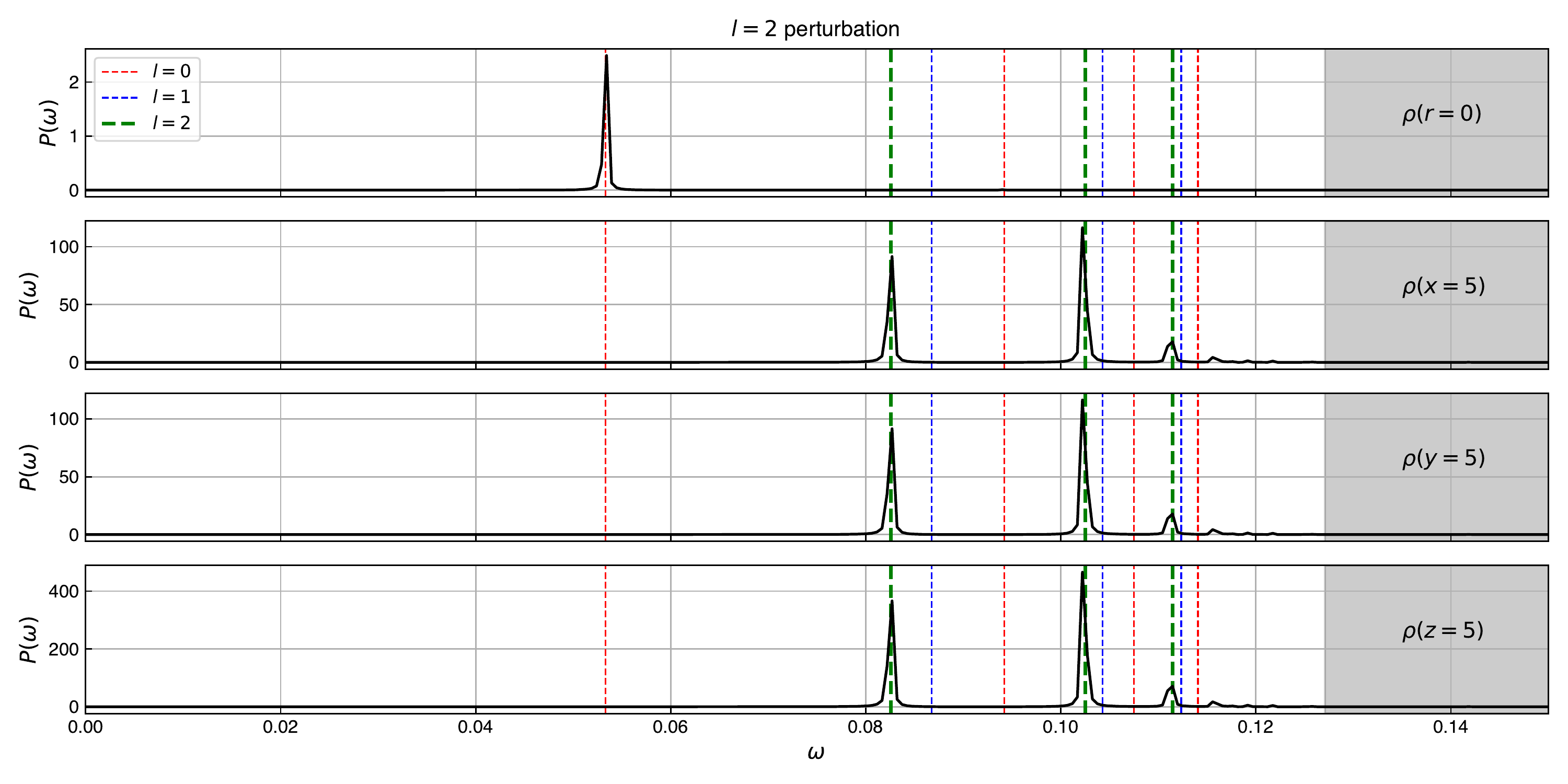}
\caption{The power spectrum of $\ell = 2 $ perturbation (black lines) at four locations: $r = 0 $, $x = 5$, $y=5$ and $z=5$.
The same color coding as \cref{fig:l0} is used. Note the difference in
vertical scale between the first and the remaining panels.
}
\label{fig:l2}
\end{center}
\end{figure}

In \cref{fig:l2}, the vertical dotted green lines correspond to the
predicted energy levels of $\ell = 2$ modes and $n = 0$ to $2$. 
There are three prominent peaks in the density power spectrum measured
away from the soliton center at the points on the $x$, $y$ and $z$
axes. They are due to the quadrupole perturbation proportional to 
$Y_{2,0} ( \theta)$. Since $Y_{2,0} (0)=-2Y_{2,0} (\pi/2)$, the power
spectrum in the $z$ direction is
four times larger than the power spectrum in the $x$ and 
$y$ directions, as observed in these panels. 
There is no resonance corresponding to any of the quadrupolar modes at
the origin since their wavefunctions vanish at $r=0$.   
Due to the mode mixing induced by the lattice, the fundamental
monopole line is present in the first panel
($r =0$) but with a much smaller power spectrum.  
\\

It is truly satisfying to see that 
the study of $\ell = 0, 1, 2$ perturbations in the wave simulation
confirms the energy eigenvalues deduced from solving the
Schr\"{o}dinger--Poisson equations.

\section{An Application: Light Soliton Evaporation}
\label{sec:scattering}

The soliton bound states play an important role in the dynamics of a
soliton immersed in the gas of axion waves \cite{Chan:2022bkz}. 
Scattering of a wave on the soliton can lead to
excitation/de-excitation of such states which reduces/increases the
number of particles in the soliton itself. In this section we use the
bound state wavefunctions found above to calculate the numerical
coefficient in the evaporation rate of light (cold) solitons
discussed in \cite{Chan:2022bkz}. It will be clear from this example
that the use of the exact wavefunctions is essential to obtain a
finite result: The Schr\"{o}dinger approximation would lead to
infrared divergences. The example will also illustrate the efficiency
of using 
the Schr\"{o}dinger approximation for the high-level and
continuum modes
to simplify the derivations.

\subsection{Review of light soliton and axion gas scattering}

We start by briefly reviewing the derivation of the light soliton
evaporation rate presented in~\cite{Chan:2022bkz}. We consider a
process when an axion from the gas ($g$) scatters on the soliton ($s$)
and kicks out another axion into the gas. The soliton evaporation rate
is given by the difference of the rates of this process and its
inverse, 
\be
\label{eq:proc}
g+ s \leftrightarrow g + g\;.
\ee
Importantly, we include into the soliton only the particles in the
ground state ($\ell=0, n=0$), whereas all excited levels, including
the discrete ones are ascribed to the gas. This division is justified
because transferring a particle from the ground state to an excited
state, even if the latter is bound, depletes the coherent condensate
in 
the soliton.  

The soliton-gas scattering is described by the non-linear term in the
axion action\footnote{As in \cref{sec:perturbations}, we 
  integrate out the Newton potential $\Phi$.} 
which contains a cubic contribution in the gas field,  
\begin{equation} 
   S_{NL} = \int dt d^3 x \left[ 
       - 2 \pi G m^2 \, |\psi |^2 \frac{1}{\Delta} |\psi |^2   
\right]  \; \supset\;
S_3=\int dt d^3 x \left[ 
- 4 \pi G m^2  \, \chi \left( \delta \psi_g  + \delta \psi^*_g  \right) 
         \frac{1}{ \Delta} |\delta \psi_g |^2  \right],
\label{eq:Sg34}
\end{equation} 
where the soliton wavefunction has the form (\ref{eq:gensol}). The gas
field here should be understood as the second-quantized operator of
perturbations in the soliton background. It is expanded in the
eigenmodes as
\be
\label{eq:psigas}
\delta\psi_g(t,{\bf x})=\sum_{\epsilon_a>0} \big(a_a\,\vf_a^{(1)}({\bf
x})\, \e^{-i\epsilon_a t}+a_a^\dagger\,{\vf_a^{(2)}}^*({\bf
x})\, \e^{i\epsilon_a t}\big)\;,
\ee
where $a_a$ and $a_a^\dagger$ are annihilation and creation operators
for an axion in the $a$th state. Note that we do not include in the
expansion the zero modes or their conjugates since they correspond to
coherent scaling of the soliton wavefunction
or motion of the soliton as a whole and thus do not
represent genuine soliton excitations. Inserting  \cref{eq:psigas}
into the action we 
obtain,
\be
\label{eq:S3as}
\begin{split}
S_3=-4\pi Gm^2\sum_{a,b,c}
2\pi\,\delta(\epsilon_a\!+\!\epsilon_b\!-\!\epsilon_c)\, 
a_a^\dagger a_b^\dagger a_c
\int d^3x \bigg[&\chi \left({\vf_a^{(1)}}^*\!\!+{\vf_a^{(2)}}^*\right)
\frac{1}{\Delta}
\left({\vf_b^{(1)}}^*\vf_c^{(1)}\!+{\vf_b^{(2)}}^*{\vf_c^{(2)}}\right)\\
&+\chi\left({\vf_c^{(1)}}+{\vf_c^{(2)}}\right)\frac{1}{\Delta}
\left({\vf_b^{(1)}}^*{\vf_a^{(2)}}^*\right)\bigg] +\text{h.c.}
\end{split}
\ee
With this action one can compute the rates of the processes
(\ref{eq:proc}) using the standard quantum-mechanical perturbation
theory. Note that the term in the second line leads to an
  $s$-channel exchange diagram. This term is generally subdominant
  compared to the contribution from the first line. For example, it is
  small if the particle $a$ belongs to the continuum since then
  its negative-frequency wavefunction $\vf^{(2)}_a$ is negligible. On the
  other hand, in the light soliton case considered here, it is
  suppressed by the hard momentum transfer, as will be clear shortly.   

We assume that the occupation numbers of the unbounded gas modes are
given by the Maxwell distribution,
\begin{equation}
      f_\k =   \frac{(4\pi)^{3/2}}{m k_g^3}\rho_g  \, \e^{-k^2/k_g^2} \, ,
   \label{fgas}
\end{equation}
where $k$ is the momentum of the mode far away from the soliton. The
case of light/cold soliton corresponds to the setup when the
characteristic gas momentum is much larger than the inverse soliton
size,  
\begin{equation}
   \nu \equiv {k_g}/{ k_s} \gg 1\, . 
\end{equation}
In this case the dominant contribution into the rates comes from the
first term in \cref{eq:S3as} and its hermitian conjugate
describing the processes $c+s\leftrightarrow a+b$
where the particles $b$ and $c$ are
`hard', i.e. have momenta of order $k_g$, but the momentum transfer is
`soft',
i.e. the difference between
their momenta, is only of order $k_s$. This `collinear' scattering is
enhanced compared to other channels due to the presence of inverse
Laplacian in the vertex which makes the amplitude inversely proportional to the
square of the momentum transfer. For hard particles of the continuous
spectrum we can neglect the negative-frequency component of the
wavefunctions and set $\vf_b^{(2)}=\vf_c^{(2)}=0$. Furthermore, their
positive-frequency components can be chosen to be simple plane waves, 
 $\vf_b^{(1)}=\e^{i{\bf k}_b{\bf x}}$, $\vf_c^{(1)}=\e^{i{\bf k}_c{\bf
     x}}$. On the other hand, we cannot drop the negative-frequency
 part for the particle $a$, which has momentum of order $k_s$ and can
 be in an unbound or bound state.

With these considerations, and including the Bose enhancement factors
for initial and final states we get the rate of the light soliton mass
change, 
\begin{equation} 
  \Gamma_s^{\rm light} \equiv \frac{1}{M_s} \frac{d M_s}{ dt}  
    = \frac{m}{2M_s} \sum_{a,b,c} 
   2\pi\, \delta(\epsilon_a+\epsilon_b-\epsilon_c)\,
   \big(f_a f_b - f_a f_c-f_b f_c\big)\,|A'_{as,bc}|^2\;.
\label{eq:NsGen}
\end{equation} 
where $f_{a,b,c}$ are the occupation numbers in the respective states
and $A'_{as,bc}$ is the matrix element following from the action
(\ref{eq:S3as}) 
stripped off the energy conservation $\delta$-function,
\begin{equation}
\label{Astrip}
   A_{as,bc}'=(4\pi Gm^2)
   \int\frac{d^3k}{(2\pi)^3k^2} V_{as}(\k) V_{bc}(-\k)\;.
\end{equation}
Here the vertex form factors are
\bseq
\label{Vs}
\begin{align} 
 &V_{as}(\k)=\int d^3x\, \left(\vf_a^{(1)}(\x)+\vf_a^{(2)}(\x)\right)
\chi(|\x|)\,\e^{i\k\x}\;,\\
&V_{bc}(\k)=\int d^3x\, \vf_b^{(1)}(\x)  \,
 {\vf_c^{(1)}}^*(\x)\,\e^{i\k\x}=(2\pi)^3\delta(\k_b-\k_c+\k)\;.
\end{align}
\eseq 
Due to the collinear scattering kinematics we have
$(f_b-f_c)/f_b=O(\nu^{-2})$ which leads to the cancellation of the
first two terms in brackets in \cref{eq:NsGen} at the leading order in
$\nu$. In this way the occupation number of the soft state $a$ drops
out of the rate. The remaining rate happens to be negative implying
that the mass of the 
soliton decreases, i.e. it evaporates. 

To proceed, it is convenient to use the characteristic momentum of
particles in the 
soliton $k_s$ as a unit to rescale other quantities:  
the coordinate, momentum, energy, and wavefunction
\begin{equation}
   \x\mapsto\boldsymbol{\xi}/{k_s}~,~~~\k\mapsto{k_s} 
      \boldsymbol{\kappa}
~,~~~\epsilon\mapsto\epsilon\, (k_s^2/m)~,~~~
\vf(\x)\mapsto k_s^{3/2}\vf(\boldsymbol{\xi})\;.
\label{eq:rescale}
\end{equation}
By converting all the variables to the rescaled ones, the scattering
rate takes the form 
\begin{equation}
   \Gamma_s^\text{light}=\frac{(4\pi
     G)^2m^3\rho_g^2}{k_g^6}\,\gamma^\text{light}_s(\nu)\;, 
\end{equation}
where
\begin{equation}
\label{gammadef}
\gamma_s^\text{light}(\nu)=-\frac{16\pi}{\mu_0}
\sum_{a} \int\frac{d^3\kappa_b\,d^3\kappa_c}{(2\pi)^3}\,
\delta\!\left(\epsilon_a+\tfrac{\kappa_b^2}{2}-\tfrac{\kappa_c^2}{2}\right)\,
\e^{-(\kappa_b^2+\kappa_c^2)/\nu^2} \,
\frac{|{\cal V}_{as}(\boldsymbol{\kappa}_b-\boldsymbol{\kappa}_c)|^2}
{|\boldsymbol{\kappa}_b-\boldsymbol{\kappa}_c|^4}\;,
\end{equation}
and the rescaled form factor is
\be
{\cal V}_{as}(\boldsymbol{\kappa})=
\int d^3\xi\,\left(\vf_a^{(1)}(\boldsymbol{\xi}) 
+\vf_a^{(2)}(\boldsymbol{\xi}) \right)\,
\chi_0(\xi)\,\e^{i\boldsymbol{\kappa\xi}}\;.
\label{dimlessVs}
\ee
All quantities in eqs.~(\ref{gammadef}), (\ref{dimlessVs}) are
dimensionless and refer to the standard soliton background
$\chi_0(\xi)$. In particular, $\mu_0$ is the standard soliton mass
introduced in 
\cref{eq:MEs}. Taking the integral over the average momentum 
$(\boldsymbol{\kappa}_b+\boldsymbol{\kappa}_c)/2$ we obtain at the
leading order in $\nu$, 
\begin{equation}
\label{F2new}
\gamma_s^\text{light}(\nu)=-C_{ls}\,\nu^2~,~~~~~~C_{ls}=\frac{8\pi^2}{\mu_0}
\sum_{a}\int \frac{d^3\kappa}{(2\pi)^3\kappa^5}|{\cal
  V}_{as}(\boldsymbol{\kappa})|^2\; .
\end{equation}
We now turn to the evaluation of the coefficient $C_{ls}$.

\subsection{Scattering rate coefficient}
 
To find $C_{ls}$, we need to sum over all the possible states of
particle $a$. 
This task may appear prohibitively complicated since
it requires knowledge of all eigenfunctions in the soliton background
entering into the form factor (\ref{dimlessVs}). We overcome this
difficulty by the following strategy. We use the exact eigenfunctions
for the modes of the few lowest levels, but the higher modes are
replaced by the Schr\"{o}dinger approximation. This allows us to
evaluate the infinite sum over states by using the sum rules for the
Schr\"{o}dinger wavefunctions. Note that it is impossible to directly apply the
completeness relations (\ref{eq:unitfull})
for the exact wavefunctions since they do not provide a sum rule for
the combination $\vf^{(1)}_a+\vf^{(2)}_a$ 
entering into the matrix
element (\ref{dimlessVs}).  
On the other hand, one 
cannot use the Schr\"{o}dinger approximation for all modes, even as a
rough estimate: We are going to see that this approach would produce
the divergence of the 
integral (\ref{F2new}) at $\kappa\to 0$ and thus would lead to a
qualitatively wrong answer. 

Upon decomposing the states into multipoles and using the spherical
symmetry of the background, we can integrate over angles which leaves
us with the expression,\footnote{In deriving this formula we have used
the decomposition of the plane wave into spherical harmonics,
\[
\e^{i\boldsymbol{\kappa\xi}}=4\pi\sum_{\ell=0}^\infty\sum_{m=-\ell}^\ell
i^\ell j_\ell(\kappa\xi)\,Y_{\ell m}(\hat{\boldsymbol{\kappa}}) \,
Y_{\ell m}^*(\hat{\boldsymbol{\xi}})\;, 
\]
where $\hat{\boldsymbol{\kappa}}$, $\hat{\boldsymbol{\xi}}$ are unit vectors in
the direction of $\boldsymbol{\kappa}$ and $\boldsymbol{\xi}$.
}  
\begin{equation}
\label{eq:Clssum}
   C_{ls} = \frac{16\pi}{\mu_0}
   \sum_{\ell n}  ( 2 \ell + 1 ) \int_0^\infty
 \frac{ d \kappa }{ \kappa^3}  \, 
      \bigg| \int_0^\infty dr \, r^2 \, 
\left( \varphi^{(1)}_{\ell n } (r)  + \varphi^{(2)}_{\ell n } (r)  \right)  
      \, \chi_0(r) \, j_\ell ( \kappa r ) \bigg|^2  \, , 
\end{equation}
where $j_\ell $ is the spherical Bessel function of the order
$\ell$. 
The sum here is taken over all excitations with strictly
positive energy $\epsilon_{\ell n}>0$. It includes discrete modes and
the states from the continuum. The monopole and dipole sectors contain
also zero modes which are not included in the sum and require special
treatment. We evaluate the contributions into 
$C_{ls}$ separately for $\ell = 0$, $\ell =1 $ and $\ell \geq 2$.

\paragraph{\bf Monopole $\ell = 0$.}

We calculate the $\ell = 0$ contribution as follows:
First, we replace the exact wavefunctions by
the Schr\"{o}dinger wavefunctions and apply the completeness relation
of the latter to obtain 
$C_{ls}^{(\rm Sch)}( \ell  = 0)$. 
Second, because for the low-lying modes the two wavefunctions differ
significantly, we consider the corrections  
to $C_{ls}^{\rm (Sch)}(\ell  = 0)$ from the four lowest modes, 
$\Delta C_{ls} ( \ell = 0, n)$, $n = 1, 2, 3, 4$. 
Hence the approximate result is  given 
by the sum of the terms, 
\begin{equation}
   C_{ls} ( \ell = 0 )  
\simeq C_{ls}^{\rm (Sch)}( \ell  = 0, n \neq 0 )  
+\sum_{n=1}^4 \Delta C_{ls} ( \ell = 0, n) \, .   
\end{equation}
Note that we do not include the 
$n =0$ state since its energy is degenerate with the soliton. 
The same procedure applies to calculating the other multipole
contributions. 

In more detail, before replacing the wavefunctions by the
Schr\"{o}dinger ones, we use the orthogonality with respect to the
inner product (\ref{eq:basisNorm1}) between the ground state $\vf_{00}$ (see
\cref{eq:groundstate}) and the excited states which implies
\begin{equation}
   \int dr\, r^2 \,   \left( \varphi^{(1)}_{0 n } (r)  +
     \varphi^{(2)}_{0 n } (r) \right) 
         \, \chi_0 (r)  = 0 \ .  \qquad n > 0  \, .
\end{equation}
It allows us to change the form of the matrix element in the
expression for $C_{ls} ( \ell = 0 )$, without changing its value, 
\begin{equation}
 \int dr \, r^2 \, \left( \varphi^{(1)}_{0 n } (r)  + \varphi^{(2)}_{0
     n } (r)  \right)  
      \, \chi_0(r) \,  j_0 ( \kappa r ) = 
   \int dr \, r^2 \, \left( \varphi^{(1)}_{0 n } (r)  +
     \varphi^{(2)}_{0 n } (r) \right )   
      \, \chi_0(r) \, \big( j_0 ( \kappa r )  - 1\big)  \, .
\end{equation}
This modification makes this matrix element manifestly of order
$O(\kappa^2)$ at small $\kappa$ and will allow us to interchange the
order of integration in $C_{ls} ( \ell = 0 )$ without encountering
infrared divergences. 

We are now ready to make the replacement
$\vf_{0n}^{(1)}+\vf_{0n}^{(2)}\mapsto \vf_{0n}^{\rm (Sch)}$ and use
the sum rules following from \cref{eq:Schrnorm},
\begin{equation}
   \sum_{n >0 } \vf^{\rm (Sch)}_{ 0 n  }  (r) \, \vf^{\rm (Sch)}_{ 0 n  }(r^\prime)    = \frac{1}{r^2} \delta ( r- r^\prime ) 
         - \frac{4 \pi} {\mu_0}  \chi_0 ( r ) \, \chi_0 (r^\prime )   , 
\end{equation}
where on the r.h.s. we have explicitly subtracted the contribution of
the ground state wavefunction $\vf_{00}^{\rm (Sch)}$ and used that it is
proportional to the soliton profile 
$\chi_0 (r) $.
Inserting this completeness relation into \cref{eq:Clssum} we obtain,
\be
\begin{split}
   C_{ls}^{\rm (Sch)}( \ell  = 0) 
      =& 
       \frac{16 \pi}{\mu_0}
    \int_0^\infty \frac{ d \kappa }{ \kappa^3}  \, 
         \int_0^\infty dr \, r^2 \,  \big(\chi_0(r)\big)^2
      \, \big( j_0 ( \kappa r )  - 1\big)^2 
       \\ 
       &-  \frac{64 \pi^2}{\mu_0^2} 
\int_0^\infty\frac{d\kappa}{\kappa^3}
\left[ \int_0^\infty dr \, r^2 \,  \big(\chi_0(r)\big)^2
      \, \big( j_0 ( \kappa r )  - 1\big) \right]^2 \, .
      \label{eq:Cl0sch}
\end{split}
\ee
It is convenient to interchange the order of integrations over
$\kappa$ and $r$, so that the integrals over $\kappa$ involving only
the Bessel functions can be performed analytically. Then the remaining
integrals over the radial coordinate are taken numerically and give
\be
\label{eq:CSch0}
C_{ls}^{\rm (Sch)}=0.772\;.
\ee

Having the numerical solutions of the exact and Schr\"{o}dinger
wavefunctions given in \cref{sec:PertTheory} 
and shown in \cref{fig:H_wave}, we evaluate
the corrections  
$\Delta C_s ( \ell = 0, n)$ for the first few modes. For a single mode,
the correction takes the form  
\begin{eqnarray}
   \Delta C_{ls} ( \ell = 0, n ) 
      &=&   C_{ls} ( \ell = 0, n )  - C_{ls}^{\rm (Sch)} ( \ell = 0, n )  
      \nonumber
      \\
       &=& \frac{16 \pi}{\mu_0}
     \int \frac{ d \kappa }{ \kappa^3}  \, 
      \left[ 
      \bigg| \int dr \, r^2 \, \left( \varphi^{(1)}_{0 n } (r)  + \varphi^{(2)}_{0 n } (r)  \right)  
      \, \chi_0(r) \, \big( j_0 ( \kappa r )  - 1\big) \bigg|^2  
      \right.
      \nonumber
      \\
      && 
      \qquad
      \qquad \,
      \quad
       - \left. \bigg| \int dr \, r^2 \,  \vf^{\rm (Sch)}_{0 n } (r)   
      \, \chi_0(r) \, \big( j_0 ( \kappa r )  - 1\big) \bigg|^2   \right],
\label{Csl0}
\end{eqnarray}
which gives numerically
\be
\Delta C_{ls}(\ell=0,n)=0.485\,,\;-0.030\,,\;-0.012\,,\;-0.006\qquad
\text{for}~~n=1,\,2,\,3,\,4\;.
\ee
Note that by far the largest correction comes from the fundamental
mode $n=1$, whereas the corrections from higher modes quickly decrease.
Adding these corrections to \cref{eq:CSch0} we will obtain the monopole
contribution $C_{ls}(\ell=0)$. 

Before presenting its numerical value it is natural to ask
about its uncertainty coming from the fact that we are still
approximating the high-level modes by the Schr\"odinger
wavefunctions. To estimate the uncertainty, we compute the relative
correction for the highest mode, 
\begin{equation}
   \delta_{\rm corr} ( \ell, n = 4 ) 
= \frac{ \Delta C_{ls}( \ell=0,n=4)  } {C_{ls}^{\rm (Sch)}( \ell=0, n=4) }\;,
\end{equation}
which happens to be about $44\%$. We expect the relative correction to
decrease for higher levels since the exact and Schr\"odinger
wavefunctions approach each other for higher frequencies.\footnote{We
  verified this trend on the low-lying modes in each multipole
  sector.} Therefore, the uncertainty is estimated as,
\begin{equation}
   \delta C_{ls} ( \ell=0 ) \lesssim 
   \delta_{\rm corr}(\ell=0,n=4) \cdot C_{ls}^{\rm (Sch)}( \ell=0,  n > 4) \, .
\end{equation}
In this way we obtain the monopole contribution with the uncertainty,
\be
\label{eq:Cls0}
C_{ls}(\ell=0)=1.21\pm0.08\;.
\ee

\paragraph{\bf Dipole $\ell = 1$.}

Similarly, the dipole contribution to $C_{ls}$ has the Schr\"{o}dinger
wavefunction summation and the exact wavefunction corrections, 
\begin{equation}
   C_{ls} ( \ell = 1 )  \simeq C_{ls}^{\rm (Sch)}( \ell  = 1, n \neq
   0)  +\sum_{n=1}^4 \Delta C_{ls} ( \ell = 1, n) \, .   
\end{equation}
where $C_{ls}^{\rm (Sch)}( \ell  = 1)$ excludes $n=0$ mode, like for
the 
monopole, since the zero mode has degenerate energy with the soliton.  
However, a naive replacement of the exact wavefunctions by the
Schr\"odinger wavefunctions leads to a spurious divergence. Indeed, in
the absence of any orthogonality relations between the dipole
Schr\"odinger modes and $\chi_0(r)$, the matrix element 
$\int dr\,r^2\,\vf_{1n}^{\rm (Sch)}(r)\chi_0(r)j_1(\kappa r)$ behaves
as $O(\kappa)$ at small $\kappa$ for any $n$. The integral over
$\kappa$ in \cref{eq:Clssum} then logarithmically diverges at the
lower end. 

The divergence is spurious and is removed by modifying the matrix
element {\em before} substituting the Schr\"odinger modes. We observe
that the exact modes with $n>0$ are orthogonal to the zero-mode
conjugate vector $\hat\vf_{10}$ (see \cref{eq:hatvf10}), implying
\begin{equation}
   \int dr r^2 \,   \left( \varphi^{(1)}_{1 n } (r)  +
     \varphi^{(2)}_{1 n } (r) \right) 
         \,  r \chi_0 (r)  = 0 \ ,  \qquad n > 0  \, .
\end{equation}
Thus, we can subtract the leading term in the Bessel function,
\be
\label{eq:Bessubtr}
j_1(\kappa r)\mapsto j_1(\kappa r)-\frac{\kappa r}{3}\;,
\ee
without changing the value of the matrix element. This leads us to
modify the form of the dipole contribution as follows:
\begin{align}
    C_{ls} ( \ell = 1 ) =
      \frac{48 \pi}{\mu_0} 
   \sum_{ n > 0 }   \int \frac{ d \kappa }{ \kappa^3}  \, 
         \bigg[&
        \int dr \, r^2 \, \left( \varphi^{(1)}_{1 n } (r)  +
          \varphi^{(2)}_{1 n } (r)  \right)   
      \, \chi_0(r) \, \left( j_1 ( \kappa r )  -  \frac{\kappa r}{3}\right)   
      \nonumber 
      \\
      &
         \times
         \int dr' \, r'^2 \, \left( \varphi^{(1)}_{1 n } (r')  +
           \varphi^{(2)}_{1 n } (r')  \right)   
      \, \chi_0(r') \,  j_1 ( \kappa r')     \bigg] \, .
\label{Csl1}
\end{align}
Note that we have made the replacement (\ref{eq:Bessubtr}) only in one
matrix element, but kept $j_1(\kappa r)$ in the other. This is
sufficient to avoid divergence at $\kappa\to 0$, and does not
introduce a new spurious divergence at $\kappa\to\infty$, which would
occur if we made the replacement  (\ref{eq:Bessubtr}) in both
factors. 

Starting from the expression (\ref{Csl1}) the rest of the analysis
proceeds in the same way as in the monopole case. Substituting the
Schr\"odinger wavefunctions and using their completeness relation we
obtain 
\be
C_{ls}^{\rm (Sch)}(\ell=1)=0.522\;.
\ee
The corrections from the exact low-lying modes are
\be
\Delta
C_{ls}(\ell=1,n)=0.119\,,\;0.019\,,\;0.006\,,\;0.002\qquad\text{for}~~
n=1,\,2,\,3,\,4\;.
\ee
The relative correction for the highest mode with $n=4$ is
$20\%$. Putting everything together we arrive at
\be
\label{eq:Cls1}
C_{ls}(\ell=1)=0.67\pm 0.07\;.
\ee
This is about a factor of two smaller that the monopole contribution
(\ref{eq:Cls0}), with a comparable uncertainty.

\paragraph{\bf Higher multipoles $\ell \geq 2 $.}

For the other multipoles, since there are no small or large-$\kappa$
divergences, we sum the contributions together with the completeness relation
without modifications, and include the corrections from the exact wavefunction, 
\begin{equation}
   C_{ls} ( \ell \geq 2 )  \simeq C_{ls}^{\rm (Sch)}( \ell   \geq  2 )  
      + \!\sum_{n=0,1,2}\Delta C_{ls} ( \ell = 2, n) 
      + \!\sum_{n=0,1}\Delta C_{ls} ( \ell = 3, n ) 
      + \Delta C_{ls} ( \ell = 4, n = 0 ) \, .
\end{equation}
We include the $n= 0$ modes here, because for $\ell \geq 2$, they
do not have the same energy as the soliton.
Using the completeness relations for the radial Schr\"{o}dinger
wavefunctions in each multipole sector and the identity 
\be
\sum_{\ell=0}^\infty (2\ell+1)\big(j_\ell(z)\big)^2=1
\ee
we obtain
\begin{equation}
    C_{ls}^{\rm (Sch)} ( \ell \geq 2 )  = 
         \frac{ 16\pi }{\mu_0}  
      \int_0^\infty \frac{ d \kappa }{ \kappa^3} \int_0^\infty\!\! dr r^2
      \big(\chi_0(r)\big)^2 \big( 
      1 - j_0^2 ( \kappa r ) - 3 j_1^2 ( \kappa r )\big) 
      =  
         \frac{ 16\pi }{ 9 \mu_0}   \,  
\int_0^\infty \!\!dr r^4 \big(\chi_0(r)\big)^2  \, .
\end{equation}
Numerical evaluation of the integral yields,
\be
C_{ls}^{\rm (Sch)}(\ell\geq 2)=2.245\;.
\ee
The exact wavefunction corrections are:
\bseq
\begin{gather}
\Delta
C_{ls}(\ell=2,n)=0.181\,,\;0.005\,,\;-0.002\qquad\text{for}~~n=0,\,1,\,2\;,\\ 
\Delta
C_{ls}(\ell=3,n)=0.003\,,\;0.001\qquad\text{for}~~n=0,\,1\;,\\
\Delta
C_{ls}(\ell=4,n=0)\sim 10^{-5}\;.  
\end{gather}
\eseq
We observe that the corrections quickly decrease with the increase of
$\ell$ and $n$. 
Still, it happens that the relative correction remains
sizeable --- of order $10\%$. Using it to estimate the uncertainty, we
obtain 
\be
\label{eq:Cls2}
C_{ls}(\ell\geq 2)=2.43\pm 0.14\;.
\ee
This is larger than the joint contribution of the monopole and
dipole. We conclude that excitation of modes with higher multipoles is
as important as that of monopole and dipole modes. Though
contributions of individual states beyond quadrupole are tiny, the
cumulative effect is significant due to a large phase space
consisting mostly of unbounded waves. 
\\ 

We now have all the multipole contributions in hand. 
Summing eqs.~(\ref{eq:Cls0}), (\ref{eq:Cls1}), (\ref{eq:Cls2}) we
obtain our final result 
\begin{equation}
   C_{ls}= 4.3 \, \pm 0.2 \, ,
\label{eq:csun}
\end{equation}
where we have added the uncertainties in quadratures. We believe this
treatment of uncertainty is
reasonable since we observed that the difference between the exact and
Schr\"odinger contributions can have different signs for different
modes. The light soliton evaporation rate found in this section is
consistent with the result of 
3D wave simulations of soliton immersed in axion gas \cite{Chan:2022bkz}.

\section{Conclusions}
\label{sec:conclusion}

We systematically studied the normal modes of boson stars. 
The negative and positive frequency perturbations are
coupled together through gravity. 
It causes significant differences between the exact normal mode 
wavefunctions and the solutions of the Schr\"{o}dinger equation
which neglects the mixing between the negative and positive
frequencies. 
One of the differences shows up in the zero modes. The exact spectrum
contains monopole and dipole zero modes, corresponding respectively 
to the $U(1)$ and
translation symmetry breaking by the soliton
background, 
whereas only a monopole zero mode is present in 
the Schr\"{o}dinger spectrum.

We derived the general properties of the soliton perturbations. 
The coupling between positive and negative frequencies leads to
non-standard orthogonality and completeness relations. 
We laid out a numerical approach to systematically find exact
wavefunction solutions 
by solving the Schr\"{o}dinger--Poisson equations. 
As far as we know, this is the first time that the complete first few
wavefunctions are presented. For higher-level modes, the  
solutions are getting closer and closer to the Schr\"{o}dinger
wavefunctions. This observation turns out very useful for evaluating
the scattering 
rates of soliton and axion gas. Further, we performed the 3D wave
simulation of soliton and its perturbations. The power
spectrum 
of the perturbations has clear resonant peaks at the frequencies of
the few first normal modes, in excellent agreement with the 
numerical wavefunction solutions.

Our results are useful for understanding the evolution of axion
solitons in various environments. 
As a concrete example, we studied the evaporation rate of a light soliton
immersed in a gas of axion waves. We demonstrated that the properties
of the exact normal mode wavefunctions are essential to obtain an
infrared finite result. We also exploited the fact that 
the Schr\"{o}dinger wavefunctions become 
a good approximation for high-level modes to evaluate an infinite sum
over final states in the expression for the rate.  
The final results are consistent
with the 3D wave simulations \cite{Chan:2022bkz}. 
It will be interesting to generalize this study to other physically
relevant situations, such as the growth of the heavy soliton by
kinetic relaxation or dynamics of fuzzy dark matter solitons in 
galaxies. 
The approach developed in this work can also be applied
to boson stars in more complicated models, such as scalar theories
with non-linear potentials
\cite{Chavanis:2011zi,Visinelli:2017ooc,2018arXiv181011473B}, 
or massive vector fields
\cite{Gorghetto:2022sue,Amin:2022pzv,Chen:2023bqy}.

\acknowledgments
We are grateful to Pierre Sikivie and Luna Zagorac
for useful discussion.
The research of J.H.-H.C. was made possible by the generosity of Eric and Wendy
Schmidt by recommendation of the Schmidt Futures program. 
The work of S.S. is supported by the Natural Sciences and Engineering
Research Council (NSERC) of Canada.  
Research at Perimeter Institute is supported in part by the Government
of Canada through the Department of Innovation,  
Science and Economic Development Canada and by the Province of Ontario
through the Ministry of Colleges and Universities.  
W.X. is supported in part by the U.S. Department of Energy under grant
DE-SC0022148 at the University of Florida.

\appendix

\section{Numerical Solution of the Eigenvalue Problem} 
\label{app:H0}

\subsection{Schr\"{o}dinger equation} 

Here we describe our numerical scheme to find the energy levels and
the eigenfunctions of \cref{eq:ODE}. The latter is supplemented by the
boundary conditions ensuring the regularity of the wavefunction at the
origin, $\vf'^{\rm (Sch)}_{\ell n}|_{r=0}=0$ if $\ell=0$, or 
$\vf^{\rm (Sch)}_{\ell n}\propto r^\ell$ if $\ell\geq 1$. Additionally, the
wavefunction is required to vanish at infinity, 
$\vf^{\rm (Sch)}_{\ell n}\to 0$
at $r\to \infty$. From now on we will omit 
the subscripts $\ell n$ on the wavefunctions and eigenvalues to reduce
clutter. 

We use two approaches. One is shooting: Imposing the initial
conditions at\footnote{In practice, we fix the conditions with a small
offset, at $r_{\rm min}\ll 1$, in order to avoid the singularity. In
more detail, we set $\vf^{\rm (Sch)}(r_{\rm min})=1$, 
$\vf'^{\rm (Sch)}(r_{\rm min})=0$ for
$\ell=0$; or $\vf^{\rm (Sch)}(r_{\rm min})=r_{\rm min}^\ell$,
$\vf'^{\rm (Sch)}(r_{\rm min})=\ell r_{\rm min}^{\ell-1}$ for
$\ell\geq 1$.} $r=0$ and choosing a trial
value of the energy $\epsilon^{\rm (Sch)}$, we solve the equation
towards some large radius $r_{\rm max}$. Then we vary the energy until
we find $\vf^{\rm (Sch)}(r_{\rm max})=0$. In the second method the
differential equation (\ref{eq:ODE}) is reduced to a system of linear
algebraic equations by discretizing the radial coordinate. In this way
the equation takes the form
\be
\label{eq:Schmatrix}
A_\ell^{\rm (Sch)}\big(\epsilon^{\rm (Sch)}\big)\, 
\upphi^{\rm (Sch)}=0\;,
\ee
where $A_\ell^{\rm (Sch)}$ is an $(N+1)\times (N+1)$ matrix and $\upphi^{\rm
  (Sch)}$ is an $(N+1)$-dimensional vector representing the discretized
wavefunction with the components
\be
\label{eq:phisdiscSch}
\upphi^{\rm(Sch)}_{I}=\vf^{\rm (Sch)}(r_I)~,~~~~r_I=I\cdot dr~,~~~~
dr=\frac{r_{\rm max}}{N}~,~~~~I=0,1,\ldots,N\;. 
\ee
We then search for the eigenvalues and eigenvectors of the matrix
equation (\ref{eq:Schmatrix}).
We have found that the two approaches give identical results. However,
it is problematic to generalize the shooting method to
higher-dimensional systems, such as the coupled SP
eqs.~(\ref{eq:speq}), whereas the matrix method is applicable also in
that case. Thus, we focus on the matrix method here. 

We use the second-order discretization scheme for the radial
derivatives. In principle, one can perform the discretization directly
in \cref{eq:ODE}. However, we prefer to obtain the discretized
Laplacian from the discretized action functional: 
This improves the accuracy of
the numerical approximation by ensuring 
that the obtained finite-difference operator is Hermitian
on the grid with the appropriate measure $r^2$. Thus we write,
\begin{align}
&\int dr\,r^2\,(\vf')^2~~\Longrightarrow~~\sum_{I=0}^{N-1}
dr\,r_{I+1/2}^2\frac{(\upphi_{I+1}-\upphi_I)^2}{dr^2}\notag\\
&=
-\sum_{I=1}^{N-1} dr\,r_I^2\,\upphi_I
\bigg[\frac{\upphi_{I+1}-2\upphi_I+\upphi_{I-1}}{dr^2}
+\frac{2}{r_I}\cdot\frac{\upphi_{I+1}-\upphi_{I-1}}{2dr}
+\frac{1}{4r_I^2} (\upphi_{I+1}-2\upphi_I+\upphi_{I-1})\bigg],
\end{align}
and read off the Laplacian from the terms in the square brackets. 
The elements of the matrix $A_\ell^{\rm
  (Sch)}$ are then set to:
\bseq
\label{eq:AlSch}
\begin{align}
&\Big[A_\ell^{\rm (Sch)}\Big]_{I,I+1}=\bigg(\frac{1}{dr^2}
+\frac{1}{r_Idr}+\frac{1}{4r_I^2}\bigg)dr^2\;,\\
&\Big[A_\ell^{\rm (Sch)}\Big]_{I,I}=\bigg(-\frac{2}{dr^2}
-\frac{1}{2r_I^2}-2\Phi_0(r_I)+2\ve_0+2\epsilon^{\rm
  (Sch)}-\frac{\ell(\ell+1)}{r_I^2}\bigg)dr^2\;,\\
&\Big[A_\ell^{\rm (Sch)}\Big]_{I,I-1}=\bigg(\frac{1}{dr^2}
-\frac{1}{r_Idr}+\frac{1}{4r_I^2}\bigg)dr^2\;,
~~~~~~~~~~~~\text{for}~~I=1,\ldots,N-1\;.
\end{align}
\eseq
Note that we have multiplied the matrix elements by $dr^2$ to make
them order-one. In addition, we set
\be
\Big[A_\ell^{\rm (Sch)}\Big]_{N,N}=1
\ee
to implement the boundary condition $\upphi^{\rm (Sch)}_N=0$,
and
\bseq
\begin{align}
&\Big[A_\ell^{\rm (Sch)}\Big]_{0,0}=-\Big[A_\ell^{\rm
  (Sch)}\Big]_{0,1}=-1\;, &&\text{for}~~\ell=0\;,\\
&\Big[A_\ell^{\rm (Sch)}\Big]_{0,0}=1\;, &&\text{for}~~\ell\geq 1\;,
\end{align}
\eseq
to implement the boundary condition at $r=0$. All other elements of
$A_\ell^{\rm (Sch)}$ are set to zero. 

Next, we find the eigenvalues of the matrix $A_\ell^{\rm (Sch)}$ by
solving the equation
\be
\label{eq:detSch}
\det\Big[ A_{\ell}^{\rm (Sch)}\big(\epsilon^{\rm (Sch)}\big)\Big]=0\;.
\ee
This gives us the numerical estimate $\epsilon_*^{\rm (Sch)}$ for the
energy level.\footnote{To check the convergence of of our procedure,
  we have found the estimates of each eigenvalue for several values of
  the grid step $dr_i$ and observed that they are well fitted by the
  quadratic dependence $\epsilon_{*i}^{\rm (Sch)}=a+b\,dr_i^2$, as it
  should be for a second-order scheme.}
Determining the eigenvectors involves a subtlety. Numerically,
\cref{eq:detSch} can never be satisfied exactly. This means that the
matrix $ A_{\ell}^{\rm (Sch)}\big(\epsilon^{\rm (Sch)}_*\big)$ is not
exactly degenerate, and \cref{eq:Schmatrix}
does not have any non-trivial solutions. To overcome this difficulty,
we use the following trick. We replace the $N$th row of the matrix by 
\be
\label{eq:AreplSch}
\Big[\tilde A_{\ell}^{\rm (Sch)}\big(\epsilon_*^{\rm
  (Sch)}\big)\Big]_{N,I}
=
\begin{cases}
\delta_{I0} &\text{for}~~\ell=0\\
\delta_{I1} &\text{for}~~\ell\geq 1
\end{cases}
\ee
where $\delta_{IJ}$ is the Kronecker symbol, and solve the equation 
\be
\label{eq:Schmatrix1}
\tilde A_{\ell}^{\rm (Sch)}\big(\epsilon_*^{\rm
  (Sch)}\big)\,\upphi^{\rm (Sch)}=B_\ell^{\rm (Sch)}~,~~~~~~~
\Big[B_\ell^{\rm (Sch)}\Big]_I=\delta_{IN}\frac{dr^\ell}{\ell!}\;.
\ee
By means of this replacement we enforce that $\upphi^{\rm (Sch)}$
satisfies $\upphi^{\rm (Sch)}_0=1$ (for $\ell=0$), or 
$\upphi^{\rm (Sch)}_1=dr^\ell/\ell!$ (for $\ell\geq 1$) and thus is
non-vanishing. This at the price of sacrificing the boundary condition
$\upphi^{\rm (Sch)}_N=0$, which we check to be satisfied with high
accuracy a posteriori. 
As the last step, we construct the interpolating functions $\vf^{\rm
  (Sch)}(r)$ using the solution $\upphi^{\rm (Sch)}_*=0$ of
\cref{eq:Schmatrix1} and normalize them by 
$\int_0^{r_{\rm max}}dr\,r^2\,|\vf^{\rm (Sch)}(r)|^2=1$. 

We have implemented the above algorithm using Mathematica
\cite{Mathematica}; the results are presented in \cref{fig:H_wave} and
\cref{table:eng}. We use the grid parameters varying from $r_{\rm
  max}=20$, $N=1000$ for the lowest modes to  
$r_{\rm max}=65$, $N=2000$ for modes with the highest $\ell$ and $n$
we consider. The precision of energy determination in all cases is
better than $10^{-4}$.

\subsection{Schr\"odinger--Poisson system} 
\label{app:SPeqsol}

The procedure described in the previous subsection can be applied with
minor modifications to the full SP system (\ref{eq:speq}). The main
difference is that now we have three functions
$(\vf^{(1)},\vf^{(2)},\delta\Phi)$, instead of one, so that the vector
$\upphi$ in \cref{eq:Schmatrix} obtained upon discretization
on a grid of length $N$ has dimension $(3N+3)$. We
order it in the following way:
\be
\label{eq:phidiscSP}
\upphi_{3I}=\vf^{(1)}(r_I)~,~~~
\upphi_{3I+1}=\vf^{(2)}(r_I)~,~~~
\upphi_{3I+2}=\delta\Phi(r_I)~,~~~~~~I=0,1,\ldots,N\;.
\ee
The matrix $A_\ell$ now has size $(3N+3)\times(3N+3)$ and is
more complicated. Its non-vanishing elements corresponding to the
inner sites of the grid with $1\leq I\leq N-1$ are:
\bseq
\begin{align}
&\big[A_\ell\big]_{3I,3I+3}=\big[A_\ell\big]_{3I+1,3I+4}=
\big[A_\ell\big]_{3I+2,3I+5}=
\bigg(\frac{1}{dr^2}+\frac{1}{r_Idr}+\frac{1}{4r_I^2}\bigg)dr^2\;,\\
&\big[A_\ell\big]_{3I,3I-3}=\big[A_\ell\big]_{3I+1,3I-2}=
\big[A_\ell\big]_{3I+2,3I-1}=
\bigg(\frac{1}{dr^2}-\frac{1}{r_Idr}+\frac{1}{4r_I^2}\bigg)dr^2\;,\\
&\big[A_\ell\big]_{3I,3I}=
\bigg(-\frac{2}{dr^2}-\frac{1}{2r_I^2}
-2\Phi_0(r_I)+2\ve_0+2\epsilon-\frac{\ell(\ell+1)}{r_I^2}
\bigg)dr^2\;,\\
&\big[A_\ell\big]_{3I+1,3I+1}=
\bigg(-\frac{2}{dr^2}-\frac{1}{2r_I^2}
-2\Phi_0(r_I)+2\ve_0-2\epsilon-\frac{\ell(\ell+1)}{r_I^2}
\bigg)dr^2\;,\\
&\big[A_\ell\big]_{3I+2,3I+2}=
\bigg(-\frac{2}{dr^2}-\frac{1}{2r_I^2}-\frac{\ell(\ell+1)}{r_I^2}
\bigg)dr^2\;,\\
&
\big[A_\ell\big]_{3I,3I+2}=
\big[A_\ell\big]_{3I+1,3I+2}=
-2\chi_0(r_I)dr^2\;,\\
&\big[A_\ell\big]_{3I+2,3I}=
\big[A_\ell\big]_{3I+2,3I+1}=-\chi_0(r_I)dr^2\;.
\end{align}
\eseq
Boundary conditions at $r=0$ are different for the monopole 
$\ell=0$ 
($\vf^{(1)\,\prime}=\vf^{(2)\,\prime}=\delta\Phi'=0$) and for higher multipoles
$\ell\geq 1$ ($\vf^{(1)}=\vf^{(2)}=\delta\Phi=0$) and are implemented
by setting
\bseq
\begin{align}
&\ell=0: ~~~~~~\big[A_\ell\big]_{0,I}=-\delta_{0,I}+\delta_{3,I}~,~~~
\big[A_\ell\big]_{1,I}=-\delta_{1,I}+\delta_{4,I}~,~~~
\big[A_\ell\big]_{2,I}=-\delta_{2,I}+\delta_{5,I}\;,\\
&\ell\geq 1: ~~~~~~\big[A_\ell\big]_{0,I}=\delta_{0,I}~,~~~~~
\big[A_\ell\big]_{1,I}=\delta_{1,I}~,~~~~~
\big[A_\ell\big]_{2,I}=\delta_{2,I}\;.
\end{align}
\eseq
Finally, setting 
\be
\label{eq:bcSPrmax}
\big[A_\ell\big]_{3N,I}=\delta_{3N,I}~,~~~~~
\big[A_\ell\big]_{3N+1,I}=\delta_{3N+1,I}~,~~~~~
\big[A_\ell\big]_{3N+2,I}=\delta_{3N+2,I}\;
\ee
imposes the vanishing boundary conditions at $r_{\rm
  max}$.\footnote{The condition $\delta\Phi(r_{\rm max})=0$ is not
  accurate for the dipole zero mode, for which $\delta\Phi$ is given
  by the derivative of the soliton gravitational potential,
  $\delta\Phi_{10}\propto\d_r\Phi_0$, and hence decreases only as
  $r^{-2}$ at large $r$. We take this into account by properly
  modifying the boundary conditions for this case.
 For all other modes
  we have found that $\delta\Phi$ decays exponentially and the
  vanishing boundary conditions at $r_{\rm max}$ provide a very good
  accuracy.}  

The next steps are the same as for the Schr\"odinger problem. We solve
the equation
\be
\label{eq:detSP}
\det\big[A_\ell(\epsilon)\big]=0
\ee
and obtain the estimate $\epsilon_*$ for the energy level. Then we
replace one of 
the boundary conditions (\ref{eq:bcSPrmax}) by the condition enforcing
that $\vf^{(1)}$ does not vanish identically,
\be
\big[A_\ell\big]_{3N,I}~~~~\to~~~~
\big[\tilde A_\ell\big]_{3N,I}=\begin{cases}
\delta_{0,I}\,,&\ell=0\\
\delta_{3,I}\,,&\ell\geq 1
\end{cases}
\ee
and find the eigenvector from the equation 
\be
\tilde A_\ell(\epsilon_*)\,\upphi=B_\ell~,~~~~~~~
\big[B_\ell\big]_I=\delta_{3N,I}\frac{dr^\ell}{\ell!}\;.
\ee
Finally, we construct the interpolating functions $\vf^{(1)}(r)$,
$\vf^{(2)}(r)$, $\delta\Phi(r)$ from the components of this vector and
normalize the solutions with $\epsilon_*>0$ according to the inner
product (\ref{eq:basisNorm1}), 
\begin{equation}
  \int_0^{r_{\rm max}} dr\, r^2\, \left( |\varphi^{(1)} (r) |^2 - |\varphi^{(2)} (r)|^2 \right)
  = 1 .  
\end{equation}
For the zero modes we instead use the normalization
$\vf^{(1)}_{00}|_{r=0}=1$, $\vf^{(1)\,\prime}_{10}|_{r=0}=1$. 
Note that for the dipole mode  $\vf_{10}$ this differs
from the normalization (\ref{eq:groundstate}) used in the analytical
calculations. 

Implementing this procedure in Mathematica \cite{Mathematica} we
obtain the results summarized in \cref{fig:H_wave} and
\cref{table:eng}. The parameters of the grid have been varied in the
range $15$ to $65$ for $r_{\rm max}$ and $800$ to $1000$ for $N$ to
reach the $O(10^{-4})$ 
precision in the determination of the energy levels.

%
%
%
%
%
%
%
%
%

\bibliography{axion}
\bibliographystyle{JHEP}

\end{document}